\documentclass[pdflatex,sn-mathphys-num,iicol]{sn-jnl}

\usepackage{anyfontsize}
\makeatletter
\renewcommand\normalsize{%
   \@setfontsize\normalsize{9pt}{10.5pt}%
}
\renewcommand\small{%
   \@setfontsize\small{9pt}{11pt}%
}
\renewcommand\footnotesize{%
   \@setfontsize\footnotesize{8pt}{10pt}%
}
\makeatother

\usepackage{graphicx}%
\usepackage{multirow}%
\usepackage{amsmath,amssymb,amsfonts}%
\usepackage{mathrsfs}%
\usepackage[title]{appendix}%
\usepackage[table]{xcolor}
\usepackage{textcomp}%
\usepackage{manyfoot}%
\usepackage{booktabs}%
\usepackage{algorithm}%
\usepackage{algorithmicx}%
\usepackage{algpseudocode}%
\usepackage{listings}%
\usepackage{mathtools}
\usepackage{braket}
\usepackage{simpler-wick}
\usepackage{slashed}
\usepackage{comment}
\usepackage{hyperref}
\usepackage{ragged2e} 
\usepackage{array}
\usepackage{tabularx}
\usepackage{bibunits}
\usepackage{tikz-cd}
\defaultbibliographystyle{sn-mathphys-num}
\defaultbibliography{bib}

\usepackage{soul}

\usepackage[numbers,sort&compress]{natbib}

\newcommand{\bb}{\boldsymbol}

\newcommand{\norm}[1]{\left\lVert #1 \right\rVert}

\begin{document}

\title{When identical particles cease to be indistinguishable: violation of statistics in quantum spacetime}


\author[1,2]{\fnm{Nicola} \sur{Bortolotti}}\email{nicola.bortolotti@cref.it}

\author[2,3]{\fnm{Catalina} \sur{Curceanu}}

\author[4,2,5]{\fnm{Antonino} \sur{Marciano}}

\author[1,2]{\fnm{Kristian} \sur{Piscicchia}}

\affil[1]{Centro Ricerche Enrico Fermi - Museo Storico della Fisica e Centro Studi e Ricerche “Enrico Fermi”, Via Panisperna 89 A, 00184, Rome, Italy}

\affil[2]{Laboratori Nazionali di Frascati, Istituto Nazionale di Fisica Nucleare, Via Enrico Fermi 54, 00044, Frascati, Italy}

\affil[3]{IFIN-HH, Institutul National pentru Fizica si Inginerie Nucleara Horia Hulubei, 30 Reactorului, 077125, M\u agurele, Romania}

\affil[4]{Center for Field Theory and Particle Physics \& Department of Physics Fudan University, Shanghai, 200433, China}

\affil[5]{INFN sezione Roma Tor Vergata, Rome, I-00133, Italy}

\abstract{
Quantum gravity may modify the fundamental symmetries that govern 
identical particles. In particular, noncommutative spacetime frameworks 
predict deformations of Bose and Fermi statistics. Here we develop a 
relativistic quantum field theory based on the most general oscillator 
algebra compatible with $\theta$-deformed Poincaré symmetry. This construction 
generalizes twisted statistics to a class of quon-like deformations 
allowing non-involutive particle exchange. We show that the resulting 
theory is consistent at both the free and interacting levels and derive 
its implications for atomic systems. Purely twisted statistics predicts 
Pauli-forbidden atomic transitions at rates incompatible with 
experiments. By contrast, a class of quon deformations suppresses such 
processes by powers of the noncommutativity scale, but only if 
superselection rules between permutation-symmetry sectors are violated. 
This implies an effective breakdown of particle indistinguishability and 
provides theoretical motivation for high-precision experimental tests of 
the Pauli exclusion principle. 
}




\maketitle

\begin{bibunit}

The indistinguishability of identical particles is one of the most profound principles of quantum physics. It underlies Bose-Einstein and Fermi-Dirac statistics and gives rise to phenomena ranging from atomic structure to superconductivity. More generally, quantum states transform under irreducible representations of the symmetric group, not necessarily restricted to the one-dimensional totally symmetric or antisymmetric ones, with superselection rules (SSRs) forbidding transitions between sectors of different permutation symmetry \cite{messiah1964symmetrization}.

The Pauli exclusion principle (PEP) emerges naturally in local quantum field theory (QFT) as a consequence of the spin-statistics theorem. The connection between spin and statistics is rooted in the structure of the net of local algebras \cite{haag1964algebraic} and follows directly from locality (i.e. local commutativity of observables) and Poincaré covariance \cite{guido1995algebraic}. However, these assumptions may be modified in quantum gravity (QG), where spacetime itself may acquire a non-classical structure.

Consistent quantum theories can nevertheless be constructed in which violations of locality and statistics are confined below arbitrarily small scales. This possibility is exemplified by the quon model \cite{Greenberg1991quon}, which interpolates between Bose, Fermi, infinite and anyonic statistics \cite{meljanac1994generalized} via a deformation of canonical commutation relations. The model preserves indistinguishability, SSRs, cluster decomposition, and CPT symmetry. However, the loss of locality obscures the status of relativistic invariance, in particular of the $S$-matrix, restricting the framework for many years to the non-relativistic regime. A weakened notion of locality for infinite statistics fields \cite{greenberg1990example} was subsequently shown to ensure Lorentz invariance of physical scattering amplitudes \cite{cao2009relativistic,cao2010self}. Contrary to earlier concerns \cite{chow2001quons}, the resulting framework admits consistent interactions involving infinite statistics particles, thereby extending the viability of the quon model to the relativistic domain.

The existence of nonlocal QFTs with modified statistics is particularly significant in QG, where sharp notions of causality fail, potentially challenging the conventional spin-statistics connection. Frameworks such as the quon model may therefore capture aspects of the low-energy limit of QG. 

On the other hand, it is widely expected that spacetime acquires a noncommutative structure at high energies, preventing arbitrarily precise localization of events \cite{doplicher1994spacetime}. A concrete realization arises for open strings ending on D-branes in a background $B$-field \cite{seiberg1999string}, where the effective low-energy theory features noncommuting coordinates satisfying $x^{\mu}\star x^{\nu} - x^{\nu}\star x^{\mu} = i\theta^{\mu\nu}$, also expressed phenomenologically as $\theta^{\mu\nu} = c^{\mu\nu}/\Lambda_\theta^2$, with $c^{\mu\nu}\sim \mathcal{O}(1)$ dimensionless coefficients and $\Lambda_\theta$ the energy scale of noncommutativity. The $\star$-product denotes the Groenewold-Moyal product \cite{groenewold1946principles,moyal1949quantum}, and these commutation relations define the Moyal spacetime.

Early field-theoretic constructions on Moyal space \cite{szabo2003quantum} faced both experimental and theoretical problems, since they explicitly broke Lorentz covariance and renormalizability was obstructed by UV-IR mixing effects in non-planar diagrams \cite{minwalla2000noncommutative}. After three decades of intense investigation a satisfactory noncommutative gauge theory remains elusive in this approach \cite{wallet2025noncommutative}. However, these theories are covariant under a quantum deformation of the Poincaré group \cite{chaichian2004lorentz,aschieri2005gravity,oeckl2000untwisting} obtained by twisting the coproduct map \cite{majid1995foundations}. This deformation modifies the action of Lorentz transformations on multiparticle states and consequently induces nontrivial $\mathcal{R}$-matrices and deformed statistics \cite{oeckl2000untwisting}. Compatibility of particle exchange with the deformed coproduct requires the twist
\begin{equation}
   f_\theta(p,q) = \exp\left(-\frac{i}{2}p\wedge_\theta q\right) , \quad p \wedge_\theta q \coloneqq p_\mu \theta^{\mu\nu} q_\nu ,
\end{equation} 
leading to a deformed algebra for the fields' creation and annihilation operators \cite{balachandran2006spin}
\begin{equation} \label{deformed algebra}
    a(\tilde p)a^\dagger(\tilde q) - \eta(p,q) f_\theta^{-2}(p,q) a^\dagger(\tilde q)a(\tilde p) = \delta(\tilde p,\tilde q), 
\end{equation}
where $\tilde p = (\bb p, \rho)$ and $\tilde q = (\bb q, \sigma)$ denote momentum and spin, $\delta(\tilde{p},\tilde{q})=\delta_{\rho\sigma}\delta^{(3)}(\bb{p}-\bb{q})$ and $\eta$ is a Lorentz-invariant function. 

Previous analyses have restricted to $\eta=\pm1$, ensuring involutive particle exchange and simplifying the formulation of QFT. In this setting, the absence of UV-IR mixing has been demonstrated \cite{oeckl2000untwisting,balachandran2006uv,bogdanovic2024braided,ciric2023braided,ciric2024braided}. Phenomenological studies \cite{addazi2018testing,piscicchia2022strongest,piscicchia2023experimental,piscicchia2023first} (see also \cite{balachandran2010non1,balachandran2010non2}) have investigated the experimental signatures of twisted statistics, highlighting the remarkable possibility of placing stringent constraints on QG scenarios from searches for PEP violation. A crucial assumption in these works is the breakdown of SSRs, allowing all atoms in a target material—independently of wavefunction symmetry—to undergo PEP-violating transitions. This enables “closed-system” experimental configurations, in contrast to the traditional “open-system” approach of Ramberg and Snow \cite{ramberg1990experimental}, which injects external electrons into the target as required for testing the quon model \cite{porcelli2025strongest,porcelli2024vip,manti2024testing}. 

However, the restriction $\eta=\pm1$ is not a fundamental requirement. Fock representations exist for general Hermitian oscillator algebras whose exchange functions take values in the complex unit disk \cite{bozejko2017fock}. Moreover, a strict braiding is necessary to regularize a QFT \cite{oeckl1999braid}. These observations motivate the exploration of the most general oscillator algebra compatible with $\theta$-deformed Poincaré symmetry, which we denote the $\mathcal{Q}_\theta$ algebra.

In this work, we formulate for the first time a QFT based on $\mathcal{Q}_\theta$-deformed fields, representing a twisted generalization of the quon fields. We demonstrate consistency at both the free and interacting levels, focusing in particular on quantum electrodynamics. We address a rigorous covariant analysis of bound states by generalizing the Bethe-Salpeter approach \cite{BetheSalpeter1951,salpeter1952mass,nakanishi1969general}, and show that atomic states inherit modified statistics. Then, we compute electromagnetic transition amplitudes that violate the PEP, both in the presence and absence of SSRs. Compatibility with experimental bounds is achieved only if SSRs are absent and for a restricted class of quon deformations, thereby excluding purely twisted statistics and implying a breakdown of indistinguishability for identical particles—a feature also observed on $\kappa$-Minkowski noncommutative spacetime \cite{arzano2023crisis,fabiano2024multi}. 

Our framework provides a relativistic generalization of the quon model and, for the first time, a physical realization linking it to QG scenarios. It establishes a consistent setting for the quantitative investigation of noncommutative QG models and provides theoretical support for closed-system experimental searches for PEP violation.

\section*{Results}\label{sec: Results}

\subsection*{Twisted quon statistics}

Unlike the twist element $f_\theta$, the function $\eta(p,q)$ cannot be determined solely from symmetry considerations. Nevertheless, it is constrained by a number of physical consistency requirements.

First, the algebra \eqref{deformed algebra} induces a positive-definite inner product on multiparticle Hilbert spaces only if $\eta(p,q)$ is real and bounded 
\begin{equation}\label{eta bounds}
-1 \leq \eta(p,q) \leq 1,
\end{equation}
implying a quon-like deformation \cite{Greenberg1991quon,meljanac1994generalized}. Indeed, Lorentz-invariant combinations of the four-momenta $p^\mu$ and $q^\mu$ can only be symmetric, so $\eta(p,q)$ must be symmetric under particle exchange. Any symmetric phase would render the norms of physical states complex (see Methods, Eq. \eqref{twisted norms}), requiring $\eta$ to be real. It follows that the exchange function
\begin{equation}
\mathcal{Q}_\theta(p,q) \coloneqq \eta(p,q)f_\theta^{-2}(p,q),
\end{equation}
is Hermitian, $\mathcal{Q}_\theta^*(p,q) = \mathcal{Q}_\theta(q,p)$, and the bound \eqref{eta bounds} then follows from the theorem proved by Bo\.{z}ejko et al. in \cite{bozejko2017fock}.

Second, at energies much lower than the noncommutativity scale $\Lambda_\theta$, which is assumed to be of order the Planck scale, experiments constrain the quon deformation for all known particles to be close to the standard bosonic or fermionic values $\eta = \pm1$, which must be recovered in the commutative limit $\Lambda_\theta\to\infty$. Assuming analyticity in the deformation, we can expand $\eta$ in powers of the dimensionless ratio $\sigma(p,q)/\Lambda_\theta$, where $\sigma(p,q)$ is a scalar function with dimensions of energy. Up to first order
\begin{equation}\label{eta expansion}
    \eta(p,q) \sim \pm \left[1 - \left(\frac{\sigma(p,q)}{\Lambda_\theta}\right)^\kappa \right], \quad \kappa>0.
\end{equation}
Different values of $\kappa$ have distinct implications for the low-energy behavior of the theory. A quon deformation, for example, induces non-vanishing field commutators at space-like separations of order $\mathcal{O}(\Lambda_\theta^{-\kappa})$. In contrast, the contribution of the twist to this violation scales as $\mathcal{O}(\Lambda_\theta^{-2})$ \cite{balachandran2007statistics}.

Crucially, when $|\eta|<1$, no algebraic relation exists between $a(\tilde p)a(\tilde q)$ and $a(\tilde q)a(\tilde p)$, or between $a^\dagger(\tilde p)a^\dagger(\tilde q)$ and $a^\dagger(\tilde q)a^\dagger(\tilde p)$. While this may appear to hinder the construction of a consistent formalism, no such relations are required to define the Fock space or to compute matrix elements of physical observables. However, this absence has profound implications: two monomials of equal degree in the creation operators that differ only by a permutation of distinct labels must be treated as independent. In the following, we will discuss how this feature affects the construction of multiparticle states.

For $n\geq2$ identical particles, the full Hilbert space is the $n$-fold tensor product $\mathscr{H}^{n} = \mathscr{H}^{\otimes n}$, where $\mathscr{H}$ denotes the one-particle Hilbert space. Permutations of particle labels generate linearly independent vectors whenever $|\eta|<1$, therefore $\mathscr{H}^n$ cannot be projected onto purely bosonic or fermionic subspaces in the conventional way. A generic vector $\ket{\psi}\in\mathscr{H}^n$ can nevertheless be decomposed into components transforming under irreducible representations of the symmetric group $S_n$
\begin{equation}\label{vector decomposition}
\ket{\psi} = \sum_\gamma c_\gamma \ket{\psi_{\gamma}} .
\end{equation}
In general, this decomposition is not unique. However, the presence of $\theta$-deformed Poincaré symmetry naturally selects a twisted action of $S_n$ on $\mathscr{H}^n$, in which the projection operators $\mathbb{P}_{\gamma_\theta}\ket{\psi} = \ket{\psi_{\gamma_\theta}}$ are constructed by deforming permutations with the twist element $f_\theta$. We denote the resulting representations by $\gamma_\theta$ to distinguish them from the untwisted ones.

The associated symmetrized states obey twisted statistics, encompassing the twisted bosons and fermions introduced in \cite{balachandran2006spin} as well as higher-dimensional mixed representations, and realizing the generalized statistics studied in \cite{liguori1995fock}. For example, the amplitudes of twisted symmetric and antisymmetric two-particle states satisfy
\begin{subequations}\label{twisted s,a components}
\begin{align}
	&\psi_{s_\theta}(\tilde p,\tilde q) = f^2_\theta(q,p) \psi_{s_\theta}(\tilde q,\tilde p), \\
	&\psi_{a_\theta}(\tilde p,\tilde q) = -f^2_\theta(q,p) \psi_{a_\theta}(\tilde q,\tilde p).
\end{align}
\end{subequations}

Each space $\mathscr{H}^n$ inherits from the $\mathcal{Q}_\theta$ algebra a deformed inner product. In the two-particle sector, it takes the explicit form
\begin{equation}\label{inner product 2 particles}
   \braket{\psi|\phi} = \int_{\tilde p,\tilde q}\psi^*(\tilde p,\tilde q)\left[\phi(\tilde  p,\tilde q) +\mathcal{Q}_\theta(q,p) \phi(\tilde q,\tilde p) \right],
\end{equation}
where $\int_{\tilde p} = \int d\bb{p}\sum_{\rho}$, which differs from the ordinary bosonic and fermionic cases by the presence of the exchange function $\mathcal{Q}_\theta$. An important consequence is that both symmetric and antisymmetric states remain normalizable. However, unlike in quon or parastatistics frameworks, untwisted symmetry sectors are not superselected, therefore transitions between states of different untwisted symmetry are in general allowed. The breakdown of these conventional SSRs originates entirely from the twist element $f_\theta(p,q)$—specifically from its antisymmetric component $\sin(p\wedge_\theta q)$—reflecting the deformation of spacetime symmetries.

When twisted symmetry is imposed, orthogonality between distinct symmetry sectors is restored, and the full Hilbert space decomposes as a direct sum of the subspaces $\mathscr{H}^n_{\gamma_\theta} = \mathbb{P}_{\gamma_\theta}\mathscr{H}^n$ carrying an irreducible representation of $S_n$. This suggests that SSRs may persist in twisted form, namely
\begin{equation}\label{twisted superselection}
    \braket{\psi_{\gamma_\theta}|O|\phi_{\gamma_\theta'}} = 0 ,
\end{equation}
for $\gamma_\theta\neq\gamma_\theta'$, provided that physical observables commute with the twisted permutation operators. As a consequence, interference between different twisted symmetry sectors becomes unobservable. Then, for a normalized state $\ket{\psi^{\mathrm n}}=\ket{\psi}/\Vert\psi\Vert$, with decomposition \eqref{vector decomposition}, the associated density operator assumes the block-diagonal form
\begin{equation}\label{block diagonal rho}
    \rho = \sum_{\gamma_\theta}|c_{\gamma_\theta}^\text{n}|^2 \ket{\psi^\text{n}_{\gamma_\theta}}\bra{\psi^\text{n}_{\gamma_\theta}} ,
\end{equation}
where $\ket{\psi^\text{n}_{\gamma_\theta}}= \ket{\psi_{\gamma_\theta}} / \Vert\psi_{\gamma_\theta}\Vert$ and the weights are given by

\begin{equation}\label{weights}
    |c_{\gamma_\theta}^\text{n}|^2 = \frac{\Vert\psi_{\gamma_\theta}\Vert^2}{\sum_{\gamma_\theta'}\Vert\psi_{\gamma_\theta'}\Vert^2} , \qquad \sum_{\gamma_\theta}|c_{\gamma_\theta}^\text{n}|^2 = 1.
\end{equation}

This structure has important physical implications. Although the amplitudes $\psi(\tilde p,\tilde q)$ and $\psi(\tilde q,\tilde p)$ generally correspond to distinct vectors in $\mathscr{H}^n$ when $|\eta|<1$, the density operator \eqref{block diagonal rho} remains invariant under particle permutations. SSRs therefore ensure indistinguishability of identical particles, despite the presence of the quon deformation and the twist.

Eq. \eqref{block diagonal rho} also clarifies the physical interpretation of multiparticle states. They correspond to incoherent mixtures of states $\ket{\psi^n_{\gamma_\theta}}$ with twisted permutation symmetry, weighted by probabilities $|c_{\gamma_\theta}|^2$ determined by the norms given in Eq. \eqref{weights} \cite{greenberg1989phenomenology}. The quon deformation $\eta$ fixes these weights (see Eq. \eqref{twisted norms} in Methods). For $\eta=\pm1$, only the twisted bosonic or fermionic sector has non-vanishing norm, and the full Hilbert space reduces to the corresponding twisted bosonic or fermionic Hilbert space. For $|\eta|<1$, all symmetry sectors contribute. In this sense, $\eta$ provides a continuous interpolation between twisted-Bose and twisted-Fermi statistics.

\begin{figure*}[t]
    \centering
    \includegraphics[width=\textwidth]{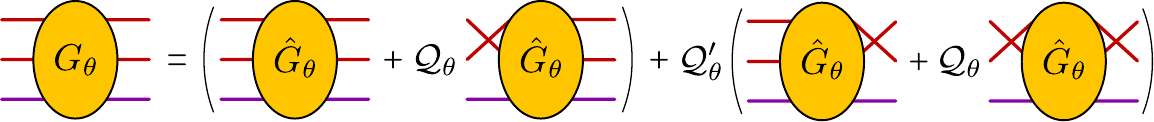}
    \caption{Six-point Green function with deformed statistics. $\mathcal{Q}_\theta$ and $\mathcal{Q}_\theta'$ are the exchange factors associated to permutations of the initial and final electrons respectively. Red and purple lines denote the electrons and the nucleus, respectively.}
    \label{fig: Q-Green function}
\end{figure*}

\subsection*{Quantum field theory}\label{subsec: free theory}

The existence of quantization schemes extending beyond Bose and Fermi statistics is well established \cite{green1953generalized}. Expanding a given field operator in terms of particle and antiparticle creation and annihilation operators (in the interacting case, within the interaction picture), the equation of motion  $i\partial_t\Psi(x) = [\Psi(x),H]$ implies
\begin{equation}\label{consistency relations}
\begin{split}
    &[a(\tilde p),H]=E_p a(\tilde p),\qquad [a^\dagger(\tilde p),H]=-E_p a^\dagger(\tilde p),\\
    &[b(\tilde p),H]=E_p b(\tilde p), \qquad \,[b^\dagger(\tilde p),H]=-E_p b^\dagger(\tilde p),
\end{split}
\end{equation}
where $E_p=\sqrt{m^2+p^2}$ and $m$ are the energy and mass of the particles and antiparticles created by $\Psi$. A central consequence of Eq. \eqref{consistency relations} is the additivity of energy, so that each particle contributes equally and linearly to total energy.  

Imposing a quadratic Hamiltonian in the creation and annihilation operators yields parastatistics of finite order \cite{green1953generalized}, including Bose and Fermi statistics as special cases. In the present construction, however, the mode expansion of the Hamiltonian must be derived consistently from the $\mathcal{Q}_\theta$ algebra itself. Specifically, one begins by expressing the Hamiltonian in Eq. \eqref{consistency relations} in terms of the $\mathcal{Q}_\theta$ field and subsequently expanding it in creation and annihilation operators. For purely twisted fields ($|\eta|=1$), the standard bilinear form of the Hamiltonian is preserved, $H = \int_{\tilde p} E_p [a^\dagger(\tilde p)a(\tilde p) + b^\dagger(\tilde p)b(\tilde p)]$. By contrast, quon deformations ($|\eta|<1$) require a modification of the Hamiltonian operator due to the failure of commutativity between the creation and annihilation operators among themselves. The resulting operator is of infinite degree, admitting an infinite expansion in the creation and annihilation operators. Up to two particle states, the number operator takes the form
\begin{equation}\label{number operator}
   N(\tilde p) = a^\dagger(\tilde p) a(\tilde p) + \int_{\tilde q} \frac{[ a^\dagger(\tilde q), a^\dagger(\tilde p)]_{\mathcal{Q}_\theta^*} \left[ a(\tilde p), a(\tilde q)\right]_{\mathcal{Q}_\theta^*}}{1-\eta^2(p,q)} + ...,
\end{equation}
where the higher-orders contribute only to matrix elements involving three or more particles and the $\mathcal{Q}_\theta$-mutator is defined as $[a(\tilde{p}),a(\tilde{q})]_{\mathcal{Q}_\theta} \coloneqq a(\tilde{p})a(\tilde{q})-\mathcal{Q}_\theta(p,q) a(\tilde{q})a(\tilde{p})$. Then $H = \int_{\tilde p} E_p [N(\tilde p) + N^c(\tilde p)]$, where an analogous expression holds for antiparticle. Eq. \eqref{number operator} agrees with the expressions given in \cite{meljanac1994generalized,stanciu1992energy}. The derivation is presented in the Supplementary Information for the specific case of a spinor field \eqref{consistency relations}.

Turning to correlation functions, Wick’s theorem extends to general braided statistics \cite{oeckl1999braid}. Any $n$-point function can be expressed in terms of two-point functions, provided that the appropriate exchange factors $\mathcal{Q}_\theta$ are included in momentum space. For instance, the four-point function $W^{(4)}(\tilde{p}_1,\tilde{p}_2,\tilde{p}_3,\tilde{p}_4) = \braket{0|\Psi(\tilde{p}_1)\Psi(\tilde{p}_2)\Psi^\dagger(\tilde{p}_3)\Psi^\dagger(\tilde{p}_4)|0}$ is given by
\begin{equation}
\begin{split}
   W^{(4)}(\tilde{p}_1,\tilde{p}_2,\tilde{p}_3,\tilde{p}_4) = \braket{0|\Psi(\tilde{p}_2)\Psi^\dagger(\tilde{p}_3)|0}\braket{0|\Psi(\tilde{p}_1)\Psi^\dagger(\tilde{p}_4)|0} \\
   + \mathcal{Q}_\theta(p_2,p_3) \braket{0|\Psi(\tilde{p}_1)\Psi^\dagger(\tilde{p}_3)|0}\braket{0|\Psi(\tilde{p}_2)\Psi^\dagger(\tilde{p}_4)|0} .
\end{split}
\end{equation}
The relative ordering of creation (annihilation) operators among themselves must be preserved. A general prescription is provided in the Supplementary Information.

Propagators retain their standard form. This holds even nonperturbatively, since the derivation of the Källén-Lehmann spectral representation relies only on one-particle states and the algebraic sector of the Poincaré group.

Finally, all correlation functions are invariant under the $\theta$-deformed Poincaré symmetry. The proof parallels that given for purely twisted fields in \cite{balachandran2007statistics} and extends straightforwardly to the quon-like case, as $\eta$ is Lorentz invariant. 

We next incorporate interactions. For the purpose of deriving PEP-violating transition amplitudes, which is the primary phenomenological objective of this work, the specific noncommutative extension of gauge symmetries is not essential. Corrections to standard amplitudes arise from both deformations of statistics and interactions, as well as their interplay. However, because PEP-forbidden processes necessarily involve modified statistics, interaction corrections do not contribute at lowest order. Nevertheless, few aspects must be addressed when a quon deformation is included.

Restricting to QED, both photon and electron creation and annihilation operators obey $\mathcal{Q}_\theta$-deformed algebras, characterized respectively by quon deformations $\eta_\gamma$ and $\eta_e$, as well as mixed photon-electron exchange relations with deformation $\eta_{\gamma e}$. The free Hamiltonians $H_\gamma$ and $H_e$ satisfy the consistency conditions \eqref{consistency relations} and generate time translations for the interaction representation fields $A_{I\mu}$ and $\Psi_I$. The time-evolution operator retains the standard Dyson form, with interaction Hamiltonian
\begin{equation}\label{interaction hamiltonian}
    H_I^\theta(\tau) = e^{i(H_\gamma+H_{e})\tau}\, H_{int}^\theta \,e^{-i(H_\gamma+H_{e})\tau} .
\end{equation}
For $|\eta|<1$, the free Hamiltonian $H_\gamma$ does not commute with the fermionic field, and similarly $H_e$ does not commute with $A_I^\mu$. Expanding Eq.~\eqref{interaction hamiltonian} in creation and annihilation operators thus yields an infinite series, inherited from the free theory. As a result, even at lowest order in the electromagnetic coupling, the interacting theory contains infinitely many terms. This is an unavoidable consequence of implementing quon deformations. As demonstrated in \cite{cao2009relativistic}, an infinite expansion of the interaction Hamiltonian is required to preserve Lorentz invariance of the $S$-matrix and conservation of statistics.

\subsection*{Atomic bound states}\label{subsec: atomic wave equations}

QFT provides the appropriate relativistic framework to describe bound states in terms of Green functions, which encode the propagation of multiparticle systems even when they do not asymptotically approach free states. The Dirac (or Breit) equation generalizes to the Bethe-Salpeter (BS) equation \cite{BetheSalpeter1951,salpeter1952mass,nakanishi1969general}, which incorporates the full relativistic interaction among the constituents.

For simplicity, we focus on the helium atom, the simplest system sensitive to violations of fermionic statistics. In the commutative theory, the relativistic “wave function” is described by the BS amplitude
\begin{equation}\label{ordinary BS amplitude}
\phi_{BS}(x_1,x_2,x_n;P) = \bra{\Omega}T\{ \Psi_e(x_1)\Psi_e(x_2)\Psi_n(x_n) \} \ket{\phi;P} ,
\end{equation}
where $\ket{\phi;P}$ denotes the bound state with total four-momentum $P^\mu$, other quantum numbers are not explicitly written. Here $\Psi_e$ and $\Psi_n$ are the Heisenberg electron and nucleus fields, and $\ket{\Omega}$ is the vacuum. By inserting a complete set of states into the associated six-point Green function $G_0^{(6)}(x_1,x_2,x_n;x_1',x_2',x_n')$ and isolating the bound state contribution, one finds
\begin{equation}\label{Green function}
    G_0(\{p_i\},\{p_i'\}) = \frac{i\phi_{BS}(\{p_i\})\phi_{BS}^{\dagger}(\{p_i'\})}{2E_P[P^0-E_P+i\epsilon]},
\end{equation}
for $P^0\to E_P$, where $\{p_i\} = p_1,p_2,p_n$. Here and in the following, we omit the superscript since we will always refer to six-point Green functions. From the perturbative expansion of $G_0$ in terms of connected Feynman diagrams, it follows that the symmetry of the Green function under permutation of the two electrons is entirely determined by the statistics of the interaction representation fields and is independent of the detailed form of the interaction Hamiltonian $\mathcal{H}_I$.

These considerations generalize to the noncommutative framework. At any perturbative order, the noncommutative Green function $G_ \theta$ contains vacuum expectation values of expressions such as $a_e(\tilde p_1)a_e(\tilde p_2) [ a_e^\dagger(\tilde k)a_e^\dagger(\tilde l) \cdot\cdot\cdot a_e(\tilde k')a_e(\tilde l') ] a_e^\dagger(\tilde p_1')a_e^\dagger(\tilde p_2')$, where operators inside the brackets originate from products of interaction Hamiltonians. The $\mathcal{Q}_\theta$ algebra then implies that $G_\theta$ is generated by deformed permutations of the external electron legs of the corresponding infinite statistics Green function $\hat{G}_\theta$, as shown in Fig. \ref{fig: Q-Green function}. 

When a quon deformation is present, the Green function loses symmetry. However, as discussed above, the state $\ket{\phi,P}$ may belong to an arbitrary twisted representation of the symmetric group, inducing the same symmetry to the BS amplitude. We therefore construct symmetrized Green functions $\mathcal{G}_{\gamma_\theta}$, with the same permutation symmetry for both the initial and final electrons, through suitable linear combinations of the Green function $G_\theta$. Using the noncommutative analogue of Eq. \eqref{Green function}, each component $G_\theta$ in the combination can be expanded in terms of BS amplitudes, yielding
\begin{equation}\label{NC Green function 1}
    \mathcal{G}_{\gamma_\theta}(\{p_i\};\{p_i'\}) \propto \frac{i\phi_{BS}^{\gamma_\theta}(\{p_i\})\phi_{BS}^{\gamma_\theta\dagger}(\{p_i'\})}{2E_P[P^0-E_P+i\epsilon]} ,
\end{equation}
where $\phi_{BS}^{\gamma_\theta}(\{p_i\}) = \hat\phi_{BS}(p_1,p_2,p_n)+s_{\gamma_\theta}f_{\theta}^2(p_2,p_1)\hat\phi_{BS}(p_2,p_1,p_n)$, $s_{\gamma_\theta}=\pm1$ and $\hat{\phi}_{BS}$ is the non-symmetric amplitude associated with $\hat{G}_\theta$. On the other hand, the perturbative expansion yields
\begin{equation}\label{NC Green function 2}
\begin{split}
    &\mathcal{G}_{\gamma_\theta}(\{p_i\};\{p_i'\}) \propto S_e^{(\gamma_\theta)}(p_1,p_2)S_n(p_n) \mathcal{I}_\theta S_n(p_n) S_e^{(\gamma_\theta)}(p_1',p_2') , \\
    &S_e^{(\gamma_\theta)}(p_1,p_2) \coloneqq S_e(p_1)S_e(p_2) + s_{\gamma_\theta}f^2_\theta(p_2,p_1) S_e(p_2)S_e(p_1) , 
\end{split}
\end{equation}
with the same overall factor as in Eq. \eqref{NC Green function 1}. $S_e$ and $S_n$ are the dressed propagators and $\mathcal{I}_\theta$ is the amputated correlation function. Eqs. \eqref{NC Green function 1} and \eqref{NC Green function 2} possess the correct symmetry properties.

 \begin{figure}[t]
	\centering
	\includegraphics[width=\linewidth]{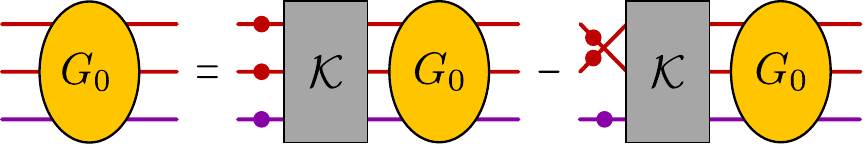}
	\caption{Structure of the six-point Green function $G_0$ for a helium-like atom in the commutative limit. The kernel $\mathcal{\mathcal{\mathcal{K}}}$ collects all irreducible two-body and three-body interactions. Red and purple lines with blobs denote dressed electron and nucleus propagators, respectively.}
	\label{fig: irreducible diagrams}
\end{figure}

In the commutative theory, one reorganizes diagrams in the amputated function $\mathcal{I}_0$ so that all two- and three-body irreducible interactions are collected into a kernel $\mathcal{K}$ (Figure \eqref{fig: irreducible diagrams}), leading to
\begin{equation}\label{BS eq}
    G_0(\{p_i\},\{p_i'\}) = S_e(p_1,p_2)S_n(p_n) (\mathcal{K} \cdot G_0)(\{p_i\},\{p_i'\}) ,    
\end{equation}
where $(A \cdot B)(\{p_i\},\{p_i'\}) = \int d^4k_1...d^4k_n A(\{p_i\},\{k_j\}) B(\{k_j\},\{p_i'\})$. Evaluating this equation at the pole yields the BS equation $\phi_{BS}(\{p_i\}) = S_e(p_1,p_2)S_n(p_n) (\mathcal{K} \cdot \phi_{BS})(\{p_i\})$. 

Assuming that in the noncommutative framework Eq. \eqref{BS eq} generalizes via $G_0 \to \mathcal{G}_{\gamma_\theta}$, then Eq. \eqref{NC Green function 1} would give 
\begin{equation}\label{3p BS equation}
    S_e^{-1}(p_1)S_e^{-1}(p_2)S_n^{-1}(p_n)\phi_{BS}^{\gamma_\theta}(\{p_i\}) = (\mathcal{K} \cdot \phi_{BS}^{\gamma_\theta})(\{p_i\}) .
\end{equation}
Time-translation invariance implies that the amplitude depends on relative energies, or relative times $x_i^0-x_j^0$ in configuration space. To obtain a single-time wave function, Eq.~\eqref{3p BS equation} must be reduced to the three-dimensional Salpeter equation \cite{salpeter1952mass}. This reduction proceeds by reconstructing the multi-time amplitude $\phi_{BS}^{\gamma_\theta}$ from a single-time wave function $\psi_{\gamma_\theta}(t,\{\bb p_i\})$ containing only positive-energy states, which is propagated to all relative time orderings through appropriate one- and two-particle propagators \cite{feldman1982relation,mittleman1989three}.

This construction can be shown to hold for generic interaction Hamiltonians when the electron field obeys pure twisted statistics. After isolating the kernel contributions and performing contractions within the kernel part and the rest of the Green function separately, one obtains for $G_\theta$ an expression proportional to $\mathcal{K}(\{p_i\},\{\tilde k_i\}) \hat{G}_\theta(\{\tilde k_i'\},\{p_i'\}) \braket{0|a(\tilde k_1) a(\tilde{k}_2)a^\dagger(\tilde k_1')a^\dagger(\tilde k_2')|0}$, and the twisted oscillator algebra produces the expected noncommutative generalization of Eq. \eqref{BS eq}.

For quon deformations, however, the validity of Eq.~\eqref{3p BS equation} is not guaranteed. Establishing it would require an explicit expression for the infinite-degree interaction Hamiltonian $\mathcal{H}_I$ \eqref{interaction hamiltonian}. If a closed equation for the twisted amplitudes $\phi_{BS}^{\gamma_\theta}$ does not exist, one can nevertheless derive a BS equation for the unsymmetrized amplitude $\phi_{BS}(\{p_i\}) = \hat\phi_{BS}(p_1,p_2,p_n) + \mathcal{Q}_\theta( p_1,p_2)\hat\phi_{BS}(p_2,p_1,p_n)$. Its reduction to the Salpeter equation then yields a wave function $\psi_{\mathcal{Q}_\theta}(\{\bb p_i\}) = \psi(\bb p_1,\bb p_2,\bb p_n) + \mathcal{Q}_\theta(p_1,p_2)\psi(\bb p_2,\bb p_1,\bb p_n)$.

\subsection*{Pauli violation with superselection rules}\label{subsec: phenomenology SSRs}

\begin{figure}
    \centering
    \includegraphics[width=\linewidth]{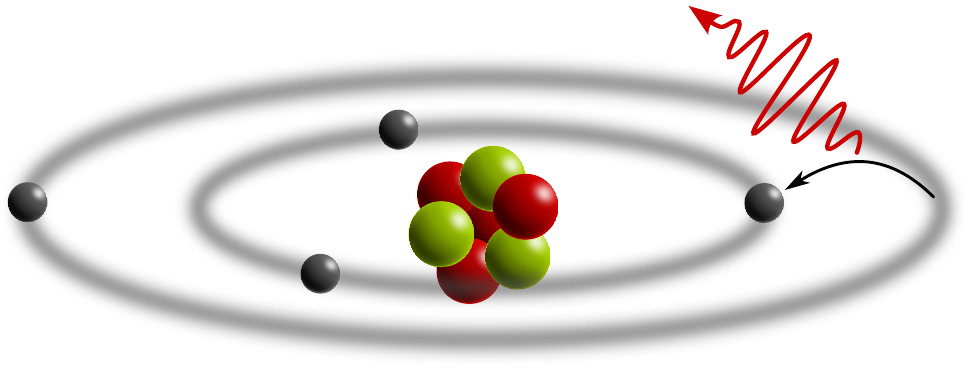}
    \caption{Pictorial representation of a radiative atomic transition between a PEP-allowed and a PEP-violating state.}
    \label{fig: PEP-violating transition}
\end{figure}

We now present the results for amplitudes and rates of PEP-violating transitions under the assumption that SSRs hold. In this case, physical states belong to superselected sectors with twisted permutation symmetry. Therefore, we focus on processes of the type 
\begin{equation}
    \ket{\psi_{\gamma_\theta}} \to \ket{\psi'_{\gamma_\theta}} \ket{\bb k \lambda} ,
\end{equation}
where a generic $n$-electron atom in state $\ket{\psi_{\gamma_\theta}}$ undergoes a radiative transition to a PEP-violating state $\ket{\psi'_{\gamma_\theta}}$ emitting a photon with momentum $\bb k$ and polarization $\lambda$. For instance, one may consider a PEP-violating state with three electrons occupying the $1s$ shell, as schematically illustrated in Figure \ref{fig: PEP-violating transition}. The photon carries a characteristic energy $\omega$, shifted with respect to PEP-allowed transitions due to the modified electronic configuration of the final state \cite{shi2018experimental,drake1989predicted}.

To first order in perturbation theory the transition rate is given by
\begin{equation}\label{general rate}
    d\Gamma = 2\pi\delta(E'-E +\omega) \left| H_{if} \right|^2 \frac{d^3k}{(2\pi)^{3}} .
\end{equation}
For helium-like atoms, the transition matrix element reads
\begin{equation}\label{PEP-violating transition matrix element}
    H_{if} = \frac{e}{\sqrt{2\omega_k}} \int d\bb{p}d\bb{q}\, D_\theta\,  \psi'^\dagger_s(\bb p - \bb k,\bb q)  \bb\alpha_{(1)}\cdot\bb\varepsilon^*_\lambda  \psi_{a_\theta}(\bb p,\bb q),
\end{equation}
where $\bb\alpha_{(1)}\cdot\bb\varepsilon^*_\lambda  \coloneqq \bb\alpha\cdot\bb\varepsilon^*_\lambda \otimes \mathbb{I}$, $\bb\varepsilon_\lambda$ is the photon polarization vector, $\psi_{a_\theta}(\bb p,\bb q)$ and $\psi_s'(\bb p,\bb q)$ are sixteen-component spinors (the latter is symmetric $\psi_s'(\bb p,\bb q)=\psi_s'(\bb q,\bb p)$) and 
\begin{equation}\label{rate deformation}
    D_\theta(p,q,k) = \frac{1-f_\theta^2(p-k,q)}{\Vert{\psi'_{a_\theta}}\Vert \Vert{\psi_{a_\theta}}\Vert} ,
\end{equation}
up to multiplicative factors arising from noncommutative corrections to the QED interaction, which do not contribute at lowest order. 

Expanding all quantities to lowest order in the noncommutativity parameter, the wave functions reduce to their ordinary forms and $\Vert \psi_{a_\theta} \Vert \sim 1$. The transition matrix elements then reduce to standard QED expressions, modified only by the norm $\Vert \psi'_{a_\theta}\Vert$ of the PEP-violating state and the twist $f_\theta^2(p,q)$. To lowest order, both contributions are linear in $\theta$ \cite{balachandran2006spin}, so the noncommutativity scale $\Lambda_\theta$ disappears from the matrix element \eqref{PEP-violating transition matrix element} and the rate. This implies that PEP-violating transitions are allowed with a probability not suppressed by $\Lambda_\theta$.

In the general case with quon deformations, SSRs imply that an atom is in one of the superselected sectors with probability given by the weights in Eq. \eqref{weights}. Consequently, an ensemble of $N$ identical atoms forms an incoherent mixture with $N_{\gamma_\theta} = N |c_{\gamma_\theta}|^2$ elements in each sector. The transition rates \eqref{general rate} must then be weighted by the fractions $N_{\gamma_\theta}$.

For electrons, experiments constrain the quon deformation to be close to $-1$, so transition rates are small perturbations of the $\eta=-1$ limit. The fraction of atoms in the twisted-symmetric sector is suppressed by some power of $\Lambda_\theta^{-1}$, depending on the leading order of $\eta$, so atoms are predominantly in the twisted-antisymmetric sector. Unlike standard quon model, which forbids PEP-violating transitions in the antisymmetric sector, such transitions may occur in all twisted sectors. Therefore, PEP-violating transitions remain unsuppressed even with a quon deformation.

\subsection*{Pauli violation for distinguishable particles}\label{subsec: phenomenology no SSRs}

We now consider the alternative scenario in which atomic states lack a definite permutation symmetry. In this case the two-particle amplitudes $\psi(p,q)=\psi_s(p,q)+\psi_a(p,q)$ and $\psi(q,p)=\psi_s(p,q)-\psi_a(p,q)$ generally correspond to two independent states, $\ket{\psi^+}$ and $\ket{\psi^-}$. Transitions are not confined to fixed symmetry sectors, implying the absence of SSRs. The corresponding density operators $\rho^\pm=\ket{\psi^\pm}\bra{\psi^\pm}$ are no longer equivalent, and interference between different sectors contributes to measurement outcomes through expectation values $\text{Tr}(\rho^\pm O)$.

PEP-forbidden transitions in helium-like atoms are then described by
\begin{equation}
    \ket{\psi^\pm} \to \ket{\psi'_s} \ket{\bb k \lambda} ,
\end{equation}
where the final electronic state has vanishing antisymmetric component. Both the symmetric and antisymmetric parts of the initial states $\ket{\psi^\pm}$ contribute to the transition amplitude, which decomposes as $H_{if}^{\pm}=H_{if}^{s}\pm H_{if}^{a}$, with
\begin{equation}\label{no SSRs matrix element}
\begin{split}
    H_{if}^\gamma = \frac{e}{\sqrt{2\omega_k}} \int d\bb{p}d\bb{q} D_\theta^\gamma\psi_{s}'^\dagger(\bb p -\bb k,\bb q)  \bb\alpha_{(1)}\cdot\bb\varepsilon^*_\lambda \psi_\gamma(\bb p,\bb q) ,
\end{split}
\end{equation}
where
\begin{equation}\label{deformation no SSRs} 
\begin{split}
    &D_\theta^s = \frac{[1+\mathcal{Q}_\theta(p,q)][1+\mathcal{Q}_\theta^*(p-k,q)] + 1 - \eta^2(p-k,q)}{\Vert{\psi'_s}\Vert\Vert{\psi}\Vert} , \\
    &D_\theta^a = \frac{[1-\mathcal{Q}_\theta(p,q)][1+\mathcal{Q}_\theta^*(p-k,q)] - 1 + \eta^2(p-k,q)}{\Vert{\psi'_s}\Vert\Vert{\psi}\Vert} .
\end{split}
\end{equation}
The explicit form of the quon deformation $\eta$ must be specified to evaluate the transition amplitudes. For atomic processes, whose energies lie far below the noncommutativity scale, $\eta$ can be expanded as in Eq.~\eqref{eta expansion}. In the nonrelativistic limit, Lorentz invariants constructed from $p^\mu$ and $q^\mu$ reduce to either $m$ or $|\mathbf p-\mathbf q|$, allowing the function $\sigma$ to be parametrized as
\begin{equation}\label{sigma non relativistic}
    \sigma_\text{nr}(p,q) = m^{1-a} |\bb p - \bb q|^a , \quad a\in\mathbb{R} .
\end{equation}
Any multiplicative coefficient can be absorbed into a redefinition of the noncommutativity scale and the coefficients $c^{\mu\nu}$, without affecting the Moyal space commutation relations. 

The PEP-violating transition rates then read
\begin{equation} \label{rate}
    \frac{d\Gamma_\text{PV}^\pm}{d\Omega} = \delta^\pm_\kappa\left(\frac{E}{\Lambda_\theta}\right) \frac{d\Gamma_0}{d\Omega} , 
\end{equation}
where $\Gamma_0$ denotes the rate of the corresponding PEP-allowed transition and $\delta^\pm_\kappa$ encodes the leading-order suppression induced by the $\mathcal{Q}_\theta$ deformation. In general, permuting the electrons in the initial state produces different rates, highlighting the breakdown of particle indistinguishability. For radiative capture processes involving a free electron, the photon angular distribution exhibits a characteristic dependence on the background-field components $c^{ij}$.

For quon deformations with $\kappa\geq4$, no suppression occurs, which is inconsistent with observations and therefore experimentally excluded. For $0<\kappa<4$, the suppression behaves as $\delta^\pm_\kappa \sim \Lambda_\theta^{-\kappa}$ for $0<\kappa\leq2$ and $\delta^\pm_\kappa \sim \Lambda_\theta^{\kappa-4}$ for $2<\kappa<4$, vanishing in the commutative limit $\Lambda_\theta \to \infty$. In particular, PEP-forbidden rates scale linearly with $\Lambda_\theta^{-1}$ for $\kappa=1,3$ and are quadratically suppressed for $\kappa = 2$. Full derivations are provided in the Supplementary Information.

\section*{Discussion}\label{sec: discussion}

Noncommutative spacetimes possess quantum symmetries that profoundly shape the structure of multiparticle states. In this context, ordinary Bose and Fermi statistics are incompatible with deformed relativistic symmetries. To preserve covariance, they must be replaced by deformed statistics satisfying $\tau_\theta\Delta_\theta = \Delta_\theta\tau_\theta$, which ensures consistency between the transposition map $\tau_\theta$ and the twisted coproduct $\Delta_\theta$ of the Poincaré Hopf algebra. Crucially, this condition fixes statistics only up to a Lorentz-invariant multiplicative function of momenta, $\eta$. To our knowledge, this freedom has not previously been explored. However, it allows statistical structures beyond the commonly studied twisted Bose and Fermi cases, with significant theoretical and phenomenological implications.

Covariant approaches based solely on twisted statistics fail to improve the UV behavior of QFT \cite{oeckl2000untwisting,balachandran2007twisted,bogdanovic2024braided,ciric2023braided,ciric2024braided}, while noncommutative gauge theories with ordinary Bose and Fermi statistics are non-renormalizable. By contrast, strict braidings—where a double particle exchange does not restore the original configuration—can regularize QFT \cite{oeckl1999braid}. Including quon deformations in the oscillator algebra therefore offers a promising framework for UV-complete theories on Moyal space and motivates exploration of their physical consequences.

In particular, we have investigated the implications for the PEP in atomic systems. Our analysis shows that any $\mathcal{Q}_\theta$-deformation modifies the permutation symmetry of atomic electrons and allows the existence of PEP-forbidden atomic states. For the case of pure twisted statistics, PEP-violating transitions occur at rates comparable to allowed transitions, far exceeding current experimental bounds. 

By contrast, a nontrivial quon deformation may be compatible with experimental data, but only at the price of violating SSRs. This entails a breakdown of indistinguishability for identical particles which manifests at higher orders in the noncommutative deformation, whereas at leading order the theory preserves ordinary statistics. A similar conclusion has been found in the Lie-algebraic noncommutative spacetime $\kappa$-Minkowski, both in the time-like \cite{arzano2023crisis} and light-like \cite{fabiano2024multi} cases. However, these scenarios are generally expected to induce excessively large violations of the PEP, in strong disagreement with experimental constraints \cite{addazi2018testing,piscicchia2023first}. For a restricted class of quon deformations with low-energy expansion characterized by exponents  $0<\kappa<4$, PEP-violating transitions become suppressed by the noncommutativity scale. 

These results provide a theoretical framework linking noncommutative quantum gravity models with high-precision atomic tests of the Pauli exclusion principle, such as those performed by the VIP and VIP-2 experiments \cite{piscicchia2022strongest,piscicchia2023experimental,piscicchia2023first,porcelli2025strongest,porcelli2024vip,manti2024testing}.

\section*{Methods}

\subsection*{Derivation of the rates with SSRs}\label{subsec: rate derivation SSRs}

Using the complete basis of free electron states, atomic states can be written as
\begin{equation}\label{general twisted state}
    \ket{\psi_{\gamma_\theta}} = \frac{1}{n!}\int_{\tilde p_1,...,\tilde p_n} \psi_{\gamma_\theta}(\tilde p_1,...,\tilde p_n) \ket{\tilde p_1,...,\tilde p_n} ,
\end{equation}
where the amplitudes $\psi_{\gamma_\theta}(\tilde p_1,...,\tilde p_n)$ are the projection of the $4^n$-component spinors $(\psi_{\gamma_\theta})_{\alpha_1,...,\alpha_n}(\bb{p}_1,...,\bb{p}_n)$ onto free Dirac spinors $u_\alpha$. For example, in the presence of a quon deformation, a two-electron atom admits both twisted symmetric and antisymmetric states, with norms
\begin{equation}\label{twisted norms}
\begin{split}
    &\norm{\psi_{s_\theta}}^2 = \int_{\tilde p, \tilde q}   \left(1+\eta(p,q)\right) |\psi_{s_\theta}(\tilde p,\tilde q)|^2   , \\
    &\norm{\psi_{a_\theta}}^2 = \int_{\tilde p, \tilde q}  \left(1-\eta(p,q)\right) |\psi_{a_\theta}(\tilde p,\tilde q)|^2   , 
\end{split}
\end{equation}
which reduce to a purely twisted antisymmetric atom for $\eta\to-1$. 

Assuming $\eta=-1$, the transition matrix element is $H_{if} = \bra{\psi'_{a_\theta},\bb k\lambda} H_I^\theta(0) \ket{\psi_{a_\theta}}$, where $H_I^\theta(t)$ is the time-dependent perturbation inducing the transition.  For helium-like atoms
\begin{equation}\label{matrix element}
    H_{if} = \int_{\tilde p,\tilde q,\tilde p',\tilde q'} \int d\bb{x}\,\frac{ \psi'^*_{a_\theta}(\tilde p',\tilde q')\braket{\tilde p',\tilde q',\tilde k|\mathcal{H}_I^\theta|\tilde p,\tilde q} \psi_{a_\theta}(\tilde p,\tilde q) }{4\norm{\psi_{a_\theta}'}\norm{\psi_{a_\theta}}} ,
\end{equation} 
where the twisted amplitudes are constructed from the full Hilbert space amplitudes $\psi(\tilde{p},\tilde{q})$ as
\begin{align}
    &\psi_{a_\theta}(\tilde{p},\tilde{q}) = \frac{1}{2}\left[ \psi(\tilde{p},\tilde{q}) - f_\theta^{-2}(p,q)\psi(\tilde{q},\tilde{p}) \right] ,\\
    &\psi'_{a_\theta}(\tilde{p},\tilde{q}) = \frac{1}{2}\psi_{s}'(\tilde p,\tilde q)(1-f^{-2}_\theta(p,q)) ,
\end{align}
with $\psi_{s}'(\tilde p, \tilde q)$ symmetric, as it describes two electrons in the same state. 

In Coulomb gauge ($\bb \nabla \cdot \bb A =0$)
\begin{equation}\label{element}
\begin{split}
\int d\bb{x} \braket{\tilde k,\tilde p',\tilde q'|\mathcal{H}_I^\theta(\bb x)|\tilde p,\tilde q} = e\int d\mu \sum_\text{pol} (2\pi)^3\delta^{(3)}(\bb l'+\bb k'-\bb l) \\
 \times \frac{F_\theta}{\sqrt{2\omega_t}} \bar u(\tilde l') \bb\alpha\cdot\bb\varepsilon^*_{\tilde k} u(\tilde l) \braket{\tilde p',\tilde q',\tilde k|\alpha^\dagger(\tilde k') a^\dagger(\tilde l') a(\tilde l) |\tilde p,\tilde q} ,
\end{split}
\end{equation}
where $d\mu=(d\bb{k}'/\sqrt{(2\pi)^32\omega_{k'}})(d\bb{l}/\sqrt{(2\pi)^32E_l})(d\bb{l}'/\sqrt{(2\pi)^32E_{l'}})$, the sum runs over electron spins and photon polarizations and $\alpha^\dagger$ is the photon creation operator. Here, $F_\theta$ accounts for noncommutative corrections to the QED interaction, eventually arising from the covariant derivative as well as from permutations of the creation and annihilation operators. Eq. \eqref{element} can be worked out using the twisted oscillator algebra
\begin{equation}\label{twisted algebra}
\begin{split}
    &a(\tilde p)a^\dagger(\tilde q) - f_\theta^{-2}(p,q) a^\dagger(\tilde q)a(\tilde p) = \delta(\tilde p,\tilde q),  \\
   &a(\tilde p)a(\tilde q) - f_\theta^2(p,q) a(\tilde q)a(\tilde p) = 0 , \\
   &a^\dagger(\tilde p)a^\dagger(\tilde q) - f_\theta^2(p,q) a^\dagger(\tilde q)a^\dagger(\tilde p) = 0 .
\end{split}
\end{equation}
Inserting the resulting expression into Eq. \eqref{matrix element} yields
\begin{equation}\label{transition matrix element}
    H_{if} = \frac{e}{\sqrt{2\omega_k}} \int_{p,q} \frac{F_\theta(k,p,q)\psi'^\dagger_{a_\theta}(\bb p-\bb k,\bb q)  \bb\alpha_{(1)}\cdot\bb\varepsilon^*_\lambda  \psi_{a_\theta}(\bb p,\bb q)}{\norm{\psi_{a_\theta}'}\norm{\psi_{a_\theta}}} ,
\end{equation}
with $\bb\alpha_{(1)}\cdot\bb\varepsilon^*_\lambda  \coloneqq \bb\alpha\cdot\bb\varepsilon^*_\lambda \otimes \mathbb{I}$. When both initial and final states have commutative limit given by standard atomic states $\psi_{a}^{(0)}(\bb{p},\bb{q})$ and $\psi_{a}'^{(0)}(\bb{p},\bb{q})$, the matrix element \eqref{transition matrix element} reduces to the ordinary QED expression. If otherwise the final state violates PEP, it reduces to Eq. \eqref{PEP-violating transition matrix element}.

The norm of the PEP-violating state is computed from $\braket{\psi_{a_\theta}'|\psi_{a_\theta}'}$ using the expansion \eqref{general twisted state} and the algebra \eqref{twisted algebra} 
\begin{equation}\label{PEP-violating norm superselection rules}
\begin{split}
    \norm{\psi_{a_\theta}'}^2 &= \int_{\tilde p,\tilde q} |\psi'_{a_\theta}(\tilde p,\tilde q)|^2\\
    &=  \int_{\tilde p,\tilde q} |\psi'_{s}(\tilde p,\tilde q)|^2 |1-f_\theta^{-2}(p,q)|^2 \\
    &= \int_{\tilde p,\tilde q} |\psi'_{s}(\tilde p,\tilde q)|^2 \left[ \sin^2(p\wedge_\theta q) + 2\sin^4\left(\frac{p \wedge_\theta q}{2}\right) \right] \\
    &\sim \int_{\tilde p,\tilde q} |\psi'_{s}(\tilde p,\tilde q)|^2  \left(\theta^{\mu\nu}p_\mu q_\mu\right)^2 ,
\end{split}
\end{equation}
which scales linearly with the deformation parameter: $\norm{\psi'_{a_\theta}}\sim\mathcal{O}(\theta)$.

\subsection*{Derivation of the rate without SSRs}

Let $\psi_{\mathcal{Q}_\theta}(\bb p,\bb q) = \psi(\bb p,\bb q) + \mathcal{Q}_\theta(p,q)\psi(\bb q,\bb p)$ denote the (time-independent) electronic component of a solution to the atomic Salpeter equation, satisfying $H\psi_{\mathcal{Q}_\theta} = E\psi_{\mathcal{Q}_\theta}$. In abstract notation, this corresponds to $\ket{\psi} = \int_{\tilde p,\tilde q} \psi(\tilde p,\tilde q)\ket{\tilde p,\tilde q}$ with $H\ket{\psi} = E\ket{\psi}$, as follows from $\bra{\tilde p,\tilde q}H\ket{\psi}=E\braket{\tilde p,\tilde q|\psi}$ and $\psi_{\alpha\beta}(\bb p,\bb q) = \sum_{sr} u_\alpha(\bb ps)u_\beta(\bb qr) \psi(\bb ps,\bb qr)$.

The matrix elements $H_{if} = \bra{\psi',\bb k\lambda} H_I^\theta(0) \ket{\psi^\pm}$ are thus evaluated using initial amplitudes $\psi(\tilde p,\tilde q)$ and $\psi(\tilde q,\tilde p)$, and final amplitude $\psi'_s(\tilde p',\tilde q')$. The interaction Hamiltonian $H_I^\theta = \exp[i(H_\gamma+H_e)] H_{int}^\theta \exp[-i(H_\gamma+H_e)] $ is of infinite degree. In undeformed QED, the leading contribution to atomic transitions is built from the transition operator $N(\tilde{p},\tilde{q}) = a^\dagger(\tilde{p}) a(\tilde{q})$, which annihilates an electron in state $\tilde{q}$ and creates one in state $\tilde{p}$. In the present framework, this structure generalizes to an infinite-degree operator defined by $[N(\tilde{p},\tilde{q}) , a^\dagger(\tilde{r})] = \delta(\tilde{q},\tilde{r}) a^\dagger(\tilde{p})$, whose explicit form follows by replacing $a^\dagger(\tilde{p}) a(\tilde{p})$ with $a^\dagger(\tilde{p}) a(\tilde{q})$ in the number operator \cite{greenberg1991interacting}. Indeed, since $H_{int}^\theta$ is linear in the photon creation operator and we consider processes with a single photon in the final state, the free photon Hamiltonian $H_\gamma$ does not contribute to $H_I^\theta$. Likewise, only the electron part of $H_e$ contributes, as the states involve no positrons. Consequently, the interaction reduces to an infinite series involving only electron operators. Moreover, while the explicit form of $H_{int}^\theta$ depends on the noncommutative extension of QED, to leading order PEP-forbidden amplitudes are controlled entirely by the modified statistics, with deformations of the covariant derivative contributing only at higher order. Accordingly, the matrix elements $\bra{\psi',\bb k\lambda} H_I^\theta(0) \ket{\psi^\pm}$ can be computed from the interaction Hamiltonian density 
\begin{equation}
    \mathcal{H}_I^\theta = \int d\mu \sum_{\text{pol}} e^{-i(\bb k+\bb p'-\bb p) \cdot \bb x}\bar{u}(\tilde{p}') \gamma^\mu \varepsilon_\mu u(\tilde{p}) \alpha^\dagger(\tilde{k}) N(\tilde{p}',\tilde{p}) ,
\end{equation}
where $d\mu=(d\bb{k}/\sqrt{(2\pi)^32\omega_k})(d\bb{p}'/\sqrt{(2\pi)^32E_{p'}})(d\bb{p}/\sqrt{(2\pi)^32E_p})$ and the sum runs over electron spins and photon polarizations. The transition operator between two-electron states takes the form
\begin{equation}\label{transition operator}
   N(\tilde p,\tilde q) = a^\dagger(\tilde p) a(\tilde q) + \int_{\tilde r} \frac{[ a^\dagger(\tilde r), a^\dagger(\tilde p)]_{\mathcal{Q}_\theta^*} \left[ a(\tilde q), a(\tilde r)\right]_{\mathcal{Q}_\theta^*}}{1-\eta^2(p,r)} .
\end{equation}
Although the commutative limit is not manifest in this expression, all matrix elements reduce to their standard values as $\theta\to0$. The evaluation of $\braket{\tilde p',\tilde q',\tilde k|\mathcal{H}_I^\theta|\tilde p,\tilde q}$ follows directly from the $\mathcal{Q}_\theta$ algebra. Using the permutation symmetry of the initial and final amplitudes then leads to Eq. \eqref{deformation no SSRs}, where the factors $[1\pm\mathcal{Q}_\theta(p,q)][1+\mathcal{Q}_\theta^*(p-k,q)]$ arise from the first term in Eq. \eqref{transition operator}, while $\pm(1 - \eta^2(p-k,q))$ originate from the second. Detailed evaluations of the matrix elements and rates are provided in the Supplementary Information.

\putbib
\end{bibunit}

\clearpage

\onecolumn
\newgeometry{left=2.2cm,right=2.2cm,top=2.6cm,bottom=2.6cm}
\fontsize{11pt}{13pt}\selectfont

\begin{center}
\Large\textbf{Supplementary Information}

\vspace{0.5cm}

\end{center}

\begin{bibunit}

\tableofcontents
\vspace{1cm}

This Supplementary Information is organized as follows. We begin by reviewing the derivation of deformed field oscillators on Moyal space, followed by an analysis of their Fock representations (Section \ref{supp: sec: Q-algebra}) and the construction of a field-theoretic framework (Section \ref{supp: sec:QFT}) for the general case of non-involutive braidings. These sections provide the necessary ingredients to discuss bound states and transition amplitudes. In Section \ref{supp: sec: atomic states}, we focus on atomic systems, presenting a field-theoretic derivation of the relativistic wave equation and the construction of atomic states with generalized statistics, explicitly demonstrating the existence of states that violate the Pauli exclusion principle. Finally, Section \ref{supp: sec: rates} presents the derivation of Pauli-violating transition rates.

\section{Covariance on the Moyal space}\label{supp: sec: covariant field theory}

Here, we recall the implementation of covariance on Moyal space, which will provide the starting point for our analysis of deformed statistics in the subsequent Sections. The existence of the twist element $\mathcal{F}_\theta$ allows one to rely on a well established approach to noncommutative field and gauge theory. In this Section, we follow the approach introduced in Ref. \cite{balachandran2006spin}. 

The isometry group of the Moyal space is the quantum group \cite{majid1995foundations} defined by the so-called $\theta$-Poincaré algebra \cite{chaichian2004lorentz}. This preserves the Poincaré algebra $\mathfrak{p}$ while allowing a nontrivial coproduct structure $\Delta_\theta : U(\mathfrak{p}) \mapsto U(\mathfrak{p}) \otimes U(\mathfrak{p})$ which deforms the universal enveloping of the Poincaré algebra $U(\mathfrak{p})$. The coproduct is defined such that the action of Lorentz transformations leaves $[x^\mu,x^\nu]=i\theta^{\mu\nu}$ invariant in all frames. This can be done by introducing the twist element 
\begin{equation}\label{supp: twist element}
    \mathcal{F}_\theta = \exp\left(-\frac{i}{2}\theta^{\mu\nu} P_\mu \otimes P_\nu\right) \in U(\mathfrak{p})\otimes U(\mathfrak{p}),
\end{equation}
which deforms the primitive coproduct $\Delta$ as
\begin{equation}\label{supp: coproduct}
    \Delta_\theta(Y) = \mathcal{F}_\theta^{-1} \Delta(Y)  \mathcal{F}_\theta .
\end{equation}

Now, let $\mathcal{A}(\mathbb{R}^{1,d})$ be the algebra of smooth functions on $\mathbb{R}^{1,d}$, equipped with the point-wise multiplication $m(\Phi \otimes \Psi)(x) = \Phi(x)\Psi(x)$ for $\Phi,\Psi\in\mathcal{A}(\mathbb{R}^{1,d})$. Later will specialize to 1+3 dimensions. The action of the twisted Poincaré group is not an automorphism of $\mathcal{A}(\mathbb{R}^{1,d})$. One is then forced to deform the multiplication map in order to satisfy the compatibility condition
\begin{equation}\label{supp: compatibility}
    m_\theta(\Delta_\theta(\Lambda) \rhd \Phi\otimes\Psi) = \Lambda \rhd m_\theta(\Phi\otimes\Psi) ,
\end{equation}
where $\Lambda$ is a suitable representation of a Lorentz transformation. Eq. \eqref{supp: compatibility} means that applying a boost to $\Phi\otimes\Psi$ and then multiplying the boosted fields is the same as applying the boost to the fields after having multiplied them. It is easy to see that the following noncommutative multiplication map

\begin{equation}\label{supp: moyal product}
    m_\theta(\Phi\otimes\Psi)(x) \coloneqq \Phi(x) \star \Psi(x) = \Phi(x) \exp\left( -\frac{i}{2} \theta^{\mu\nu} \overleftarrow{\partial_\mu} \,\overrightarrow{\partial}_\nu \right) \Psi(x),
\end{equation}
called the \textit{Moyal $\star$-product}, satisfies Eq. \eqref{supp: compatibility}. 

Note that the operation of deforming $\mathcal{A}(\mathbb{R}^{1,d})$ into the algebra $\mathcal{A}_\theta(\mathbb{R}^{1,d})$ equipped with the noncommutative multiplication map \eqref{supp: moyal product}, makes at the same time spacetime coordinates noncommutative. Indeed, the application of the $\star$-product to the coordinate functions yields the canonical commutation relations
\begin{equation}
    x^\mu \star x^\nu - x^\nu \star x^\mu = i \theta^{\mu\nu} .
\end{equation}
This is analogous to the Groenewold-Moyal phase space formulation of QM \cite{groenewold1946principles,moyal1949quantum} where, unlike the approach to phase space quantization $\{x,p\}=1\to[\hat x, \hat p]=i\hbar$, phase space retains a smooth structure and the algebra of the observable is deformed into a noncommutative algebra equipped by a $\star$-product. The two formulations are equivalent and connected by the Wigner-Weyl transform.

Let us consider the action of twisted Lorentz transformations on tensor products of real scalar free fields (the argument extends to arbitrary tensor fields). This action is immediately given by the coproduct \eqref{supp: coproduct} 
\begin{equation}\label{supp: lorentz on tensor products}
\begin{split}
    \Delta_\theta(\Lambda)\, \Phi \otimes \Psi &= \int d\mu(p)d\mu(q) \,\Phi(p) \Psi(q)\,\Delta_\theta(\Lambda) e_p \otimes e_q \\
    &= \int d\mu(p)d\mu(q) \,\Phi(p) \Psi(q)\,f_\theta^{-1}(\Lambda p,\Lambda q) f_\theta(p,q) e_{\Lambda p} \otimes e_{\Lambda q} \\
    &= \int d\mu(p)d\mu(q) \,\Phi(\Lambda^{-1}p) \Psi(\Lambda^{-1}q)\,f_\theta(\Lambda^{-1}p,\Lambda^{-1}q) f^{-1}_\theta(p,q) e_{p} \otimes e_{q} ,
\end{split}
\end{equation}
where $d\mu(p)$ is a Lorentz-invariant measure, 
\begin{equation}
	e_p(x) = e^{-i p \cdot x} \in \mathcal{A}_\theta(\mathbb{R}^{1,d}) ,	
\end{equation} 
and $f_\theta$ is defined by
\begin{equation}
    f_\theta(p,q) = \exp\left(-\frac{i}{2}p\wedge_\theta q\right) , \quad p \wedge_\theta q \coloneqq p_\mu \theta^{\mu\nu} q_\nu .
\end{equation}
Therefore, we can deduce the action of twisted Lorentz transformations on the Fourier amplitudes to be
\begin{equation}
	 \Phi(p)\Psi(q) \xrightarrow{\Lambda} f_\theta(\Lambda^{-1}p,\Lambda^{-1}q) f^{-1}_\theta(p,q) \Phi(\Lambda^{-1}p)\Psi(\Lambda^{-1}q) .
\end{equation}

We are now ready to proceed to quantize the fields. They admit the usual expansion in terms of positive and negative frequency modes  
\begin{equation}
    \Phi(x) = \Phi^+(x) + \Phi^-(x) = \frac{1}{(2\pi)^{3/2}}\int\frac{d^dp}{\sqrt{2E_p}} \left( c(\bb p) e^{-ip\cdot x} + c^\dagger(\bb p) e^{ip\cdot x} \right) ,
\end{equation}
where $c(\bb p)$ and $c(\bb p)^\dagger$ are quantum operators. As for the classical Fourier amplitudes, these operators transform as 
\begin{equation}
    c(\bb p) c^\dagger(\bb q) \xrightarrow{\Lambda} f_\theta(\Lambda^{-1}p,\Lambda^{-1}q) f_\theta^{-1}(p,q) c(\Lambda^{-1}\bb p)c^\dagger(\Lambda^{-1}\bb q),
\end{equation}
etc., so the canonical commutation relations (CCRs) cannot be compatible with twisted Lorentz transformations. It turns out that there exists a generalization of the CCRs in terms of a complex-valued exchange factor $\mathcal{Q}_\theta(p,q)$
\begin{equation}\label{supp: Q algebra}
    [c(\bb p),c^\dagger(\bb q)]_{\mathcal{Q}_\theta} \coloneqq c(\bb p)c(\bb q)^\dagger - \mathcal{Q}_\theta(p,q) c(\bb q)^\dagger c(\bb p) = \delta^{(d)}(\bb p - \bb q) ,
\end{equation}
which preserves covariance under the twisted Poincaré group. This is given by those $\mathcal{Q}_\theta$ which make commutate the following diagram
\begin{equation}
\begin{tikzcd}
c(\bb p)c^\dagger(\bb q) \arrow[r,"\Lambda"] \arrow[d, "\tau_\theta"]  & \Delta_\theta(\Lambda)\,c(\bb p)c^\dagger(\bb q)\arrow[r,"\tau_\theta"] & \tau_\theta\Delta_\theta(\Lambda)\,c(\bb p)c^\dagger(\bb q)\arrow[ddll,dashed, leftrightarrow] \\
\tau_\theta\, c(\bb p)c^\dagger(\bb q) \arrow[d,"\Lambda"]  \\
\Delta_\theta(\Lambda)\tau_\theta\,c(\bb p)c^\dagger(\bb q) 
\end{tikzcd}
\end{equation}
where $\tau_\theta$ is the transposition or flip operator $\tau_\theta c(\bb p)c^\dagger(\bb q) = \mathcal{Q}_\theta(p,q)c^\dagger(\bb q)c(\bb p)$. This leads to the condition 
\begin{equation}
    \frac{\mathcal{Q}_\theta(\Lambda^{-1}p,\Lambda^{-1}q)}{\mathcal{Q}_\theta(p,q)} = \frac{f_\theta^{-2}(\Lambda^{-1}p,\Lambda^{-1}q)}{f_\theta^{-2}(p,q)} ,
\end{equation}
whose solution is
\begin{equation}\label{supp: Q function}
    \mathcal{Q}_\theta(p,q) = \eta(p,q) f_\theta^{-2}(p,q),
\end{equation}
where $\eta$ is a Lorentz invariant function. 

The same argument can also be applied to the commutators $[c(\bb p),c(\bb q)]_{\mathcal{Q}_\theta'}$ and $[c^\dagger(\bb p),c^\dagger(\bb q)]_{\mathcal{Q}_\theta''}$, leading to the following exchange functions  
\begin{equation}\label{supp: Q'&Q''}
    \mathcal{Q}'_\theta(p,q) = \eta'(p,q) f_\theta^{2}(p,q) , \quad \mathcal{Q}''_\theta(p,q) = \eta''(p,q) f_\theta^{2}(p,q) ,
\end{equation}
where $\eta'$ and $\eta''$ are Lorentz invariant functions of the momenta, in principle different from $\eta$ and from each other. However, for $\eta\neq\pm1$ these relations are not consistent. Indeed, computing the vacuum matrix element $\bra{0} c(\bb q') c(\bb p') c^\dagger(\bb p) c^\dagger(\bb q) \ket{0}$ using only the mixed relation \eqref{supp: Q algebra} yields
\begin{equation}\label{supp: two free particle normalization}
    \delta^{(d)}(\bb p -\bb p')\delta^{(d)}(\bb q -\bb q') + \eta(p, q) f_\theta^2(\bb q,\bb p) \delta^{(d)}(\bb p -\bb q')\delta^{(d)}(\bb q -\bb  p'),
\end{equation}
whereas permuting the particle order with $[c(\bb p),c(\bb q)]_{\mathcal{Q'}}$ and $[c^\dagger(\bb p),c^\dagger(\bb q)]_{\mathcal{Q''}}$ produces modified expressions, which agree only for $\eta=\eta'=\eta''=\pm1$. To derive Eq. \eqref{supp: two free particle normalization}, we have used the vacuum condition $c(\bb p)\ket{0} = 0$.

To conclude, for particles with a generic spin the twisted quon algebra writes
\begin{equation}\label{supp: Q algebra new}
	a(\tilde p)a^\dagger(\tilde q) - \mathcal{Q}_\theta(p,q)a^\dagger(\tilde q)a(\tilde p) = \delta(\tilde p,\tilde q),
\end{equation}
where $\tilde{p} = (\bb p,\rho)$ denotes momentum and spin projection, and $\delta(\tilde{p},\tilde{q})=\delta_{\rho\sigma}\delta^{(d)}(\bb p - \bb q)$.

\section{Multiparticle state spaces}\label{supp: sec: Q-algebra}

The algebra \eqref{supp: Q algebra new} induces deformed inner products on the multiparticle state spaces, determined by writing the generic $n$-particle states $\ket{\psi},\ket{\phi}\in\mathscr{H}^{n}$ as
\begin{subequations}
\begin{align}
    &\ket{\psi} = \int_{\tilde p_1,...,\tilde p_n}\psi(\tilde p_1,...,\tilde p_n) a^\dagger(\tilde p_1) \cdot\cdot\cdot a^\dagger(\tilde p_n)\ket{0} ,\\
    &\ket{\phi} = \int_{\tilde p_1,...,\tilde p_n}\phi(\tilde p_1,...,\tilde p_n) a^\dagger(\tilde p_1) \cdot\cdot\cdot a^\dagger(\tilde p_n) \ket{0} ,
\end{align}
\end{subequations}
where $\int_{\tilde{p}} = \int d^3p \sum_\rho$, and evaluating $\braket{\psi|\phi}$ using Eq. \eqref{supp: Q algebra new} and the vacuum condition $a(\tilde{p})\ket{0}=0$ for all momenta and spins. We introduce the mapping $\mathcal{T}_\theta : \pi_i \mapsto \mathcal{T}_\theta^i$  from the adjacent permutations $\pi_i\in S_n$ to the operators $\{\mathcal{T}_\theta^i:i=1,...,n-1\}$ acting on $\mathscr{H}^{n}$ as 
\begin{equation}\label{supp: non involutive flip operators}
    \mathcal{T}_\theta^i \phi(\tilde p_1,...,\tilde p_i,\tilde p_{i+1},...,\tilde p_n) = \mathcal{Q}_\theta(p_i,p_{i+1}) \phi(\tilde p_1,...,\tilde p_{i+1},\tilde p_i,...,\tilde p_n) .
\end{equation}
The mapping $\mathcal{T}_\theta$ extends multiplicatively to arbitrary elements of the symmetric group $S_n$. The operators $\mathcal{T}_\theta^i$ are self-adjoint and satisfy the braid relations
\begin{subequations}
\begin{align}
    \mathcal{T}_\theta^i \mathcal{T}_\theta^j &= \mathcal{T}_\theta^j \mathcal{T}_\theta^i, \quad |i-j|\geq2, \\
    \mathcal{T}_\theta^i\mathcal{T}_\theta^{i+1}\mathcal{T}_\theta^i &= \mathcal{T}_\theta^{i+1} \mathcal{T}_\theta^i \mathcal{T}_\theta^{i+1} .
\end{align}    
\end{subequations}
They are not, however, involutions
\begin{equation}
	(\mathcal{T}_\theta^i)^2 = \eta^2(p_i,p_{i+1}) \neq 1 ,
\end{equation}
unless $\eta=\pm1$. We further introduce the operator $T_\theta$ on $\mathscr{H}^{n}$ defined by
\begin{equation}
    T_\theta = \frac{1}{n!} \sum_{\pi\in S_n} \mathcal{T}_\theta(\pi) .
\end{equation}
Then, the modified inner product on $\mathscr{H}^{n}$ is given by the matrix elements of $T_\theta$ 
\begin{equation}\label{supp: inner product}
        \braket{\psi|\phi} \coloneqq \braket{\psi|T_\theta|\phi}_0 ,
\end{equation}
where $\braket{\cdot|\cdot}_0$ denotes the undeformed inner product of the full Hilbert space $\mathscr{H}^{n}$. For example, $\braket{\varphi_1\otimes \varphi_2|\xi_1\otimes \xi_2}_0 = \braket{\varphi_1|\xi_1}\braket{\varphi_2|\xi_2}$. For all $n$, the operators $T_\theta$ are self-adjoint and positive \cite{bozejko2017fock}, ensuring that $\braket{\cdot|\cdot}$ are well-defined.

It is convenient to focus on two-particle states. In this case, the inner product reduces to
\begin{equation}\label{supp: inner product 2 particles}
        \braket{\psi|\phi} = \int_{\tilde p,\tilde q}\psi^*(\tilde p,\tilde q)\left[\phi(\tilde  p,\tilde q) +\mathcal{Q}_\theta(q,p) \phi(\tilde q,\tilde p) \right].
\end{equation}
An important implication of the $\mathcal{Q}_\theta$ algebra is that states with different symmetries 
\begin{subequations}\label{supp: s,a components}
\begin{align}
	\psi_{s}(\tilde p,\tilde q) = \psi(\tilde p,\tilde q) + \psi(\tilde q,\tilde p) ,\\
	\psi_{a}(\tilde p,\tilde q) = \psi(\tilde p,\tilde q) - \psi(\tilde q,\tilde p)  ,
\end{align}
\end{subequations}
are no longer orthogonal. Their inner product is given by
\begin{equation}
\begin{split}
    \braket{\psi_s|\phi_a} &= \int_{\tilde p,\tilde q}\psi_s^*(\tilde p,\tilde q)\left[\phi_a(\tilde  p,\tilde q) +\mathcal{Q}_\theta(q,p) \phi_a(\tilde q,\tilde p) \right] \\
	&= \int_{\tilde p,\tilde q}\psi_s^*(\tilde p,\tilde q)\phi_a(\tilde  p,\tilde q)\left[1 -\mathcal{Q}_\theta(q,p) \right] \\
	&= -\int_{\tilde p,\tilde q}\psi_s^*(\tilde p,\tilde q)\phi_a(\tilde  p,\tilde q)\mathcal{Q}_\theta(q,p).
\end{split}
\end{equation}
Using the explicit expression $\mathcal{Q}_\theta = \eta f_\theta^{-2}$ and the symmetry properties of the amplitudes $\psi_{s,a}$
\begin{equation}
\begin{split}
    -\int_{\tilde p,\tilde q} \psi_s^*(\tilde p,\tilde q)\phi_a(\tilde  p,\tilde q)\mathcal{Q}_\theta(p,q) &= -\int_{\tilde p,\tilde q} \psi_s^*(\tilde p,\tilde q)\phi_a(\tilde  p,\tilde q)\eta(p,q) [\cos(p\wedge_\theta q) -i\sin(p \wedge_\theta q)] \\
    &= \int_{\tilde p,\tilde q} \psi_s^*(\tilde q,\tilde p)\phi_a(\tilde  q,\tilde p)\eta(p,q) [\cos(p\wedge_\theta q) -i\sin(p \wedge_\theta q)] ;
\end{split}
\end{equation}
then exploiting the symmetry properties of $\eta$, $\cos(p \wedge_\theta q)$, $\sin(p \wedge_\theta q)$ and of the integration measure, one obtains

\begin{equation}\label{supp: untwisted states inner product}
	\begin{split}
		-\int_{\tilde p,\tilde q} \psi_s^*(\tilde p,\tilde q)\phi_a(\tilde  p,\tilde q)\mathcal{Q}_\theta(p,q) &= \int_{\tilde q,\tilde p} \psi_s^*(\tilde p,\tilde q)\phi_a(\tilde  q,\tilde p)\eta(q,p) [\cos(q\wedge_\theta p) -i\sin(q \wedge_\theta p)]\\
		&= \int_{\tilde q,\tilde p} \psi_s^*(\tilde p,\tilde q)\phi_a(\tilde  q,\tilde p)\eta(p,q) [\cos(p\wedge_\theta q) +i\sin(p \wedge_\theta q)]\\
		&= \int_{\tilde p,\tilde q} \psi_s^*(\tilde p,\tilde q)\phi_a(\tilde  q,\tilde p)\eta(p,q) [\cos(p\wedge_\theta q) +i\sin(p \wedge_\theta q)] .
	\end{split}
\end{equation}
The term proportional to $\cos(p\wedge_\theta q)$ cancels since it is an odd function of $\tilde p$ and $\tilde q$, whereas the other does not vanish due to $\sin(p\wedge_\theta q)$, which yields
\begin{equation}
    \braket{\psi_s|\phi_a} = i\int_{\tilde p,\tilde q}\psi_s^*(\tilde p,\tilde q)\phi_a(\tilde  p,\tilde q) \sin(p\wedge_\theta q) .
\end{equation}
This result shows that transitions between symmetric and antisymmetric states are allowed
\begin{equation}
    \braket{\psi_s|O|\phi_a} \neq 0 ,
\end{equation} 
therefore the subspaces $\mathscr{H}_s^{2}$ and $\mathscr{H}_a^{2}$ are no longer superselected on the Moyal space.

The proper irreducible decomposition of the states in the full Hilbert space $\mathscr{H}^{2}$ involves twisted symmetric and antisymmetric components 
\begin{subequations}\label{supp: twisted s,a components}
\begin{align}
	\psi_{s_\theta}(\tilde p,\tilde q) \coloneq \psi(\tilde p,\tilde q) + f_\theta^{-2}(p,q)\psi(\tilde q,\tilde p) , \\
	\psi_{a_\theta}(\tilde p,\tilde q) \coloneq \psi(\tilde p,\tilde q) - f_\theta^{-2}(p,q)\psi(\tilde q,\tilde p) ,
\end{align}
\end{subequations}
which satisfy
\begin{equation}\label{supp: twisted symmetry}
	\psi_{\gamma_\theta}(\tilde p,\tilde q) = \pm f_\theta^2(p,q) \psi_{\gamma_\theta}(\tilde q, \tilde p) ,
\end{equation}
where the $+$ ($-$) sign corresponds to twisted-symmetric (antisymmetric) states. Unlike in the limiting cases $\eta=\pm1$, where only one of these sectors survives, both are normalizable for $|\eta|<1$. The corresponding norms $\norm{\psi_{\gamma_\theta}} \coloneqq \sqrt{\braket{\psi_{\gamma_\theta}|\psi_{\gamma_\theta}}}$ are given by
\begin{subequations}\label{supp: twisted norms}
\begin{align}
    &\norm{\psi_{s_\theta}}^2 = \int_{\tilde p, \tilde q}   \left(1+\eta(p,q)\right) |\psi_{s_\theta}(\tilde p,\tilde q)|^2   , \label{supp: norm twisted-anti} \\
    &\norm{\psi_{a_\theta}}^2 = \int_{\tilde p, \tilde q}  \left(1-\eta(p,q)\right) |\psi_{a_\theta}(\tilde p,\tilde q)|^2   . \label{supp: norm twisted-sym} 
\end{align}
\end{subequations}

Evaluating the inner product between states with different twisted symmetries, one finds that they are orthogonal. Indeed
\begin{equation}
\begin{split}
    \braket{\psi_{s_\theta}|\phi_{a_\theta}} &= \int_{\tilde p,\tilde q}\psi_{s_\theta}^*(\tilde p,\tilde q)\left[\phi_{a_\theta}(\tilde  p,\tilde q) +\mathcal{Q}_\theta(q,p) \phi_{a_\theta}(\tilde q,\tilde p) \right] \\
	&= \int_{\tilde p,\tilde q}\psi_{s_\theta}^*(\tilde p,\tilde q)\phi_{a_\theta}(\tilde  p,\tilde q)\left[1 -\eta(p,q) \right] ,
\end{split}
\end{equation}
where in the second line we used Eq. \eqref{supp: twisted symmetry}, and using the symmetry properties of all the terms in the integrand, one finds

\begin{equation}\label{supp: orthogonality}
	\begin{split}
		\braket{\psi_{s_\theta}|\phi_{a_\theta}} &= \int_{\tilde p,\tilde q}\psi_{s_\theta}^*(\tilde p,\tilde q)\left[\phi_{a_\theta}(\tilde  p,\tilde q) +\mathcal{Q}_\theta(q,p) \phi_{a_\theta}(\tilde q,\tilde p) \right] \\
		&= \int_{\tilde p,\tilde q}\psi_{s_\theta}^*(\tilde p,\tilde q)\phi_{a_\theta}(\tilde  p,\tilde q)\left[1 -\eta(p,q) \right] \\
		&= \int_{\tilde p,\tilde q} f^{-2}_\theta(q,p) \psi_{s_\theta}^*(\tilde q,\tilde p)\phi_{a_\theta}(\tilde  p,\tilde q)\left[1 -\eta(p,q) \right] \\
		&= -\int_{\tilde p,\tilde q} f^{-2}_\theta(q,p) \psi_{s_\theta}^*(\tilde q,\tilde p) f^{2}_\theta(q,p) \phi_{a_\theta}(\tilde  q,\tilde p)\left[1 -\eta(p,q) \right] \\
		&= -\int_{\tilde q,\tilde p} f^{-2}_\theta(q,p) \psi_{s_\theta}^*(\tilde q,\tilde p) f^{2}_\theta(q,p) \phi_{a_\theta}(\tilde  q,\tilde p)\left[1 -\eta(q,p) \right] \\
		&=  -\braket{\psi_{s_\theta}|\phi_{a_\theta}}  ,
	\end{split}
\end{equation}
which implies that the inner product vanishes. The full Hilbert space $\mathscr{H}^2$ is then the direct sum of the twisted bosonic and fermionic subspaces obtained from the projection operators  
\begin{subequations}
    \begin{align}
        &\mathbb{P}_{s_\theta} = \frac{1}{2!} \sum_{\pi\in S_2} \mathcal{P}_\theta(\pi) ,   \\
        &\mathbb{P}_{a_\theta} = \frac{1}{2!} \sum_{\pi\in S_2} \text{sgn}(\pi)\mathcal{P}_\theta(\pi) ,
    \end{align}
\end{subequations}
where $\mathcal{P}_\theta(\pi)$ are permutations on $\mathscr{H}^2$ constructed by twisting the flip operator as $\tau \to \mathcal{F}_\theta^{-2} \tau$. For arbitrary number of particles and Young tableau $\gamma$, one can construct the projection operator $\mathbb{P}_{\gamma_\theta}$ as usual using the Young symmetrizer and the twisted permutations
\begin{equation}\label{supp: general projection operator}
    \mathbb{P}_{\gamma_\theta} \propto \sum_{\substack{\pi_r\in R_\gamma\\
  \pi_c\in C_\gamma}} \text{sgn}(\pi_c) \mathcal{P}_\theta(\pi_r\pi_c) ,
\end{equation}
where $R_\gamma$ and $C_\gamma$ denote the subgroups of $S_n$ that preserve, respectively, each row and each column of $\gamma$. 

This result suggests that superselection rules may persist in a generalized form, applying to states with twisted symmetry. In this case, physical observables correspond to Hermitian operators which commute with the twisted permutations $\mathcal{P}_\theta$. Each eigenvector of $O$ thus belongs to an irreducible subspace $\mathscr{H}^n_{\gamma_\theta} = \mathbb{P}_{\gamma_\theta}\mathscr{H}^n$, and the spectral decomposition $O = \sum_{n,\gamma_\theta} \ket{o_n^{\gamma_\theta}}O_n^{\gamma_\theta} \bra{o_n^{\gamma_\theta}}$, together with the orthogonality of states with different permutation symmetries, yields the superselection rule
\begin{equation}\label{supp: twisted superselection}
    \braket{\psi_{s_\theta}|O|\phi_{a_\theta}} = 0 .
\end{equation}

Remarkably, this implies that interference between states belonging to different twisted sectors is unobservable. To see this, we derive the density operator associated to the normalized vector $\ket{\psi^\text{n}} = \ket{\psi}/\norm{\psi}$. Using Eq. \eqref{supp: twisted s,a components} and $\norm{\psi}^2 = \norm{\psi_a}^2+\norm{\psi_s}^2$, the latter can be expressed as 
\begin{equation}
    \psi^\text{n}(\tilde p, \tilde q) = c_{a_\theta}\, \psi_{a_\theta}^\text{n}(\tilde p, \tilde q) + c_{s_\theta}\, \psi_{s_\theta}^\text{n}(\tilde p, \tilde q) ,
\end{equation}
with coefficients  
\begin{equation}\label{supp: alpha beta}
    c_{a_\theta} = \frac{\norm{\psi_{a_\theta}}}{\sqrt{\norm{\psi_{a_\theta}}^2+\norm{\psi_{s_\theta}}^2}} , \quad c_{s_\theta} = \frac{\norm{\psi_{s_\theta}}}{\sqrt{\norm{\psi_{a_\theta}}^2+\norm{\psi_{s_\theta}}^2}} ,
\end{equation}
satisfying $|c_{a_\theta}|^2+|c_{s_\theta}|^2=1$. The density operator is therefore
\begin{equation}\label{supp: non-diagonal density operator}
\begin{split}
    \rho = &|c_{a_\theta}|^2 \ket{\psi^\text{n}_{a_\theta}}\bra{\psi^\text{n}_{a_\theta}} + c_{a_\theta}c_{s_\theta}  \ket{\psi^\text{n}_{a_\theta}}\bra{\psi^\text{n}_{s_\theta}}\\ 
    &+ c_{s_\theta} c_{a_\theta} \ket{\psi^\text{n}_{s_\theta}}\bra{\psi^\text{n}_{a_\theta}} + |c_{s_\theta}|^2 \ket{\psi^\text{n}_{s_\theta}}\bra{\psi^\text{n}_{s_\theta}} .
\end{split} 
\end{equation}
Due to the superselection rule \eqref{supp: twisted superselection}, the off-diagonal elements of $\rho$ do not contribute to the result of any physical measurement, as can be seen by evaluation of the trace $\text{tr}(\rho O)$. Therefore, the density operator \eqref{supp: non-diagonal density operator} can be equivalently written in the block-diagonal form
\begin{equation}\label{supp: block diagonal density operator}
    \rho = |c_{a_\theta}|^2 \ket{\psi^\text{n}_{a_\theta}}\bra{\psi^\text{n}_{a_\theta}} + |c_{s_\theta}|^2 \ket{\psi^\text{n}_{s_\theta}}\bra{\psi^\text{n}_{s_\theta}} .
\end{equation}
This result has deep physical implications. Note that $\psi^\text{n}(\tilde p, \tilde q)$ and $\psi^\text{n}(\tilde q, \tilde p)$ represent distinct states in the full Hilbert space: $\psi^\text{n}(\tilde q, \tilde p) = -c_{a_\theta}\, \psi_{a_\theta}^\text{n}(\tilde p, \tilde q) + c_{s_\theta}\, \psi_{s_\theta}^\text{n}(\tilde p, \tilde q)\neq \psi^{n}(\tilde p, \tilde q)$. Nevertheless, the expression \eqref{supp: block diagonal density operator} is invariant under permutation of the two particles, as can be verified from the matrix elements $\braket{\tilde p,\tilde q|\rho|\tilde p,\tilde q}$. Thus, the SSRs \eqref{supp: twisted superselection} guarantees indistinguishability of identical particles, even though this may appear, at first glance, to conflict with a quon deformation. 

We conclude this section by showing why the restriction $-1\leq\eta\leq1$ is necessary.  Consider the general expression $\eta(p,q) = |\eta(p,q)| e^{i\vartheta(p,q)}$, which modifies the norms \eqref{supp: norm twisted-anti} and \eqref{supp: norm twisted-sym} to
\begin{subequations}\label{supp: complex-valued norms}
\begin{align}
    &\norm{\psi_{a_\theta}}^2 = \int_{\tilde p, \tilde q} \left[1+|\eta(p,q)|(\cos\vartheta(p,q) + i\sin \vartheta(p,q))\right] |\psi_{a_\theta}(\tilde p,\tilde q)|^2   , \\
    &\norm{\psi_{s_\theta}}^2 = \int_{\tilde p, \tilde q} \left[1-|\eta(p,q)|(\cos\vartheta(p,q) + i\sin \vartheta(p,q))\right] |\psi_{s_\theta}(\tilde p,\tilde q)|^2   .
\end{align}
\end{subequations}
These expressions are inevitably complex unless $\vartheta = 0$. Furthermore, assuming $\eta(p,q)>1$ in some region $\Omega$, it would be possible to choose a state in $\mathscr{H}^{2}_{s_\theta}$ with support in $\Omega$ such that Eq.  \eqref{supp: norm twisted-sym} would reduce to
\begin{equation}
    \norm{\psi_{s_\theta}^{(\Omega)}}^2 = \int_\Omega \left(1-\eta(p,q)\right) |\psi_{s_\theta}^{(\Omega)}(\tilde p,\tilde q)|^2 <0.
\end{equation}
Analogously, $\eta(p,q)<1$ would imply twisted-antisymmetric states with negative squared norms.

\section{Quantum field theory}\label{supp: sec:QFT}

Here, we construct QFT for $\mathcal{Q}_\theta$-deformed fields. To simplify the analysis, we restrict to purely spatial noncommutativity ($c^{0i}=0$), thereby avoiding potential issues with unitarity.

\subsection{Consistency of the quantization scheme}\label{supp: sec: consistency}

Here we show that free fermionic fields satisfy the canonical equations of motion 
\begin{equation}\label{supp: canonical equation}
    i\partial_t\Psi(x) = [\Psi(x),H].
\end{equation}
Expanding the fermionic field $\Psi$ in terms of $\mathcal{Q}_\theta$-deformed oscillators this leads to the relations
\begin{equation}\label{supp: consistency relations}
\begin{split}
    &[a_s(\bb p),H]=E_p a_s(\bb p),\qquad [a_s^\dagger(\bb p),H]=-E_p a_s^\dagger(\bb p),\\
    &[b_s(\bb p),H]=E_p b_s(\bb p), \qquad \,[b_s^\dagger(\bb p),H]=-E_p b_s^\dagger(\bb p),
\end{split}
\end{equation}
where $E_p=\sqrt{m^2+p^2}$ and $m$ are the energy and mass of the particles and antiparticles created by $\Psi$. The antiparticle operators satisfy exactly the same algebra \eqref{supp: Q algebra new}. On the Moyal space, the free Hamiltonian retains the same form as in the commutative theory\footnote{We recall that a free theory has the same action of the associated commutative theory, since the action is quadratic in the fields and the Moyal product is closed, namely
\begin{equation}
    \int d^4x f(x) \star g(x) = \int d^4x g(x) \star f(x) = \int d^4x f(x) g(x) .
\end{equation}
This property can be proven by using the asymptotic expansion of the Moyal product and repeatedly integrating by parts.}. The Hamiltonian is given by
\begin{equation}\label{supp: dirac hamiltonian}
\begin{split}
    H &= \int d^3x \Psi^\dagger(x) h_D \Psi(x)\\
    &= \int d^3x \left[ \Psi^{-\dagger}(x) h_D \Psi^+(x) - \mathcal{\bar Q}_\theta \Psi^-(x) h_D\Psi^{+\dagger}(x)\right] ,
\end{split}
\end{equation}
where $h_D = -i\bb\alpha\cdot\bb{\nabla} + m\beta$ is the one-particle Dirac Hamiltonian, $\Psi^\pm$ are the positive and negative frequency components of the field, namely $\Psi^+(x) = \sum_s\int (d^3p/\sqrt{(2\pi)^32E_p}) e^{-ip\cdot x}a(\tilde p)$ and $\Psi^-(x) = \sum_s\int (d^3p/\sqrt{(2\pi)^32E_p}) v(\tilde p) e^{ip\cdot x}b^\dagger(\tilde p)$. We defined $\mathcal{\bar Q}_\theta \coloneqq \mathcal{Q}_\theta(p,p)=\eta(p,p)$, due to the anti-symmetry of $\theta^{\mu\nu}$ which yields $\exp(ip_\mu\theta^{\mu\nu}p_\nu)=1$. To obtain the second line in Eq. \eqref{supp: dirac hamiltonian}, we used the $\mathcal{Q}_\theta$-mutator of the antiparticle operators evaluated at $\bb p=\bb q$, that is $b_s(\bb p) b_r^\dagger(\bb p)=-\mathcal{\bar Q}_\theta b_r^\dagger(\bb p)b_s(\bb p) + \delta_{sr}\delta^{(3)}(0)$, and omitted the infinite constant term given by $\delta^{(3)}(0)$. Inserting the expression \eqref{supp: dirac hamiltonian} into the commutator $[\Psi(x),H]$, with some algebraic manipulations, yields
 \begin{align}\label{supp: [psi,H]}
    &h_D\Psi^+(x) = [\Psi^+(x),H] + \int d^3y \Psi^{-\dagger}(y) h_D^y [ \Psi^+(y), \Psi^+(x) ]_{\mathcal{Q}_\theta^*} , \\
    &\mathcal{\bar Q}_\theta h_D\Psi^-(x) = [\Psi^-(x),H] + \mathcal{\bar Q}_\theta \int d^3y  h_D^y [ \Psi^-(x), \Psi^-(y) ]_{\mathcal{Q}_\theta^*} \Psi^{+\dagger}(y), \label{supp: bar Q}
\end{align}
where $h_D^y$ acts on $\Psi^\pm(y)$, and we defined the $\mathcal{Q}_\theta$-mutators as  
\begin{equation}\label{supp: measure}
    [\Psi^+(x), \Psi^+(y)]_{\mathcal{Q}_\theta^*} \coloneqq \frac{1}{(2\pi)^3}\sum_{sr} \int \frac{d^3p}{\sqrt{2E_p}}\frac{d^3q}{\sqrt{2E_q}} e^{-ip \cdot x} e^{-i q \cdot y} u(\tilde p) u(\tilde q) \left[a(\tilde p), a(\tilde p) \right]_{\mathcal{Q}_\theta^*}.
\end{equation}
In the commutative limit $\Lambda_\theta\to\infty$, $\mathcal{Q}_\theta=1$ and the integrals in Eq. \eqref{supp: [psi,H]} vanish via standard canonical anti-commutation relations, yielding $[\Psi,H]=h_D\Psi$. For $\Lambda_\theta$ finite, the $\mathcal{Q}_\theta$-mutators $[\Psi^\pm(x),\Psi^\pm(y)]_{\mathcal{Q}_\theta^*}$ still vanish if the operators satisfy strict twisted statistics ($\eta=-1$); in this case Eq. \eqref{supp: canonical equation} holds with the bilinear Hamiltonian $H{\vert_{\eta=1}} = \int_{\tilde p} E_p [a^\dagger(\tilde p)a(\tilde p) + b^\dagger(\tilde p)b(\tilde p)]$. For quon-like deformations $|\eta|<1$, however, the $\mathcal{Q}_\theta$-mutators $[\Psi^\pm(x),\Psi^\pm(y)]_{\mathcal{Q}_\theta^*}$ do not vanish. To investigate this, we move to momentum space and assume the validity of relations \eqref{supp: consistency relations}. This leads to the following relations
\begin{equation}
\begin{split}
    h_D(\bb p) u_s(\bb p) a_s(\bb p) =& E_p  u_s(\bb p)a_s(\bb p) \\
    &+ \sum_{r,t}\int d^3q a_r^\dagger(\bb q) \left[ a_t(\bb q), a_s(\bb p)  \right]_{\mathcal{Q}_\theta^*} \left(\frac{1}{2E_q} u_r^\dagger(\bb q) h_D(\bb q) u_t(\bb q) \right) u_s(\bb p), \label{supp: h_D 1}
\end{split}
\end{equation}
and 
\begin{equation}
\begin{split}
    h_D(-\bb p) v_s(\bb p) b_s^\dagger(\bb p) =& - \mathcal{\bar Q}_\theta^{-1} E_p  v_s(\bb p) b_s^\dagger(\bb p) \\
    &+ \sum_{r,t}\int d^3q \left[ b_t^\dagger(\bb q), b_s^\dagger(\bb p)  \right]_{\mathcal{Q}_\theta^*} b_r(\bb q) \left(\frac{1}{2E_q} v_r^\dagger(\bb q) h_D(-\bb q) v_t(\bb q) \right) v_s(\bb p) .\label{supp: h_D 2}
\end{split}
\end{equation}
As long as $a_s(\bb p)a_r(\bb q)$ and $b_s^\dagger(\bb p)b_r^\dagger(\bb q)$ have non-vanishing $\mathcal{Q}_\theta$-mutators, these equations can be solved iteratively. After rather long but straightforward calculations, we find
\begin{equation}\label{supp: h_D recursive}
\begin{split}
     &h_D(\bb p) u_s(\bb p) a_s(\bb p) = E_p  u_s(\bb p)a_s(\bb p) \\
     &+\sum_{r}\int d^3q \frac{u_s(\bb p) a_r^\dagger(\bb q)}{1-\eta^2(p,q)} \left[ (E_q+E_p\eta^2(p,q)) a_r(\bb q) a_s(\bb p) + \mathcal{Q}_\theta(p,q)(E_p+E_q)a_s(\bb p) a_r(\bb q) \right]  \\
     &+\sum_{r,t,v}\int d^3qd^3k[\cdot\cdot\cdot]\left(\frac{1}{2E_k} u_t^\dagger(\bb k) h_D(\bb k) u_v(\bb k) \right) u_s(\bb p),
\end{split}
\end{equation}
and an analogous equation for antiparticles. The terms in square brackets $[\cdot\cdot\cdot]$ in the last line contain two creation operators and three annihilation operators, so that the final line contributes only to matrix elements which involve three or more particles. The factor $(1-\eta^2)^{-1}$ in the second line arises from a geometric series $\sum_{k=0}^\infty [\eta(p,q)]^k$, which converges due to the constraint $|\eta(p,q)|<1$ for all values of the momenta. Inserting the results into Eq. \eqref{supp: dirac hamiltonian}, one obtains the Hamiltonian as an infinite series in creation and annihilation operators. Alternatively, by a rearrangement of terms, we may express the Hamiltonian in the following fashion 
\begin{equation}
    H = \int d^3p E_p \left[ N(\bb p) + N^c(\bb p) \right] ,
\end{equation}
where now $N$ and $N^c$ are particle and antiparticle number operators of infinite rank. At the end, the explicit expression for the particle number operator, up to annihilation of all one- and two-particles, is the following 
\begin{equation}\label{supp: Q number operator}
    N(\bb p) = \sum_s a_s^\dagger(\bb p) a_s(\bb p) + \sum_{s,r}\int d^3q \frac{1}{1-\eta^2(p,q)} \left[ a_r^\dagger(\bb q), a_s^\dagger(\bb p)\right]_{\mathcal{Q}_\theta^*} \left[ a_s(\bb p), a_r(\bb q)\right]_{\mathcal{Q}_\theta^*} + ... ,
\end{equation}
with a corresponding expression for $N^c(\bb p)$.

\subsection{Correlation functions}

We proceed by showing that the generic $n$-point correlation function can be computed in terms of two-point correlation functions using a generalization of the Wick's theorem. This result is consistent with other analyses \cite{Greenberg1991quon,oeckl1999braid,ciric2023braided,bogdanovic2024braided}. We focus on a single scalar field, but the conclusions are generalizable to collection of fields of arbitrary spin.

Wick's theorem states that a given string of creation and annihilation operators can be rewritten as the normal-ordered product of the string, plus the normal-ordered product after all single contractions among operator pairs, plus all double contractions, and so on until full contractions. The normal-ordering is defined as usual with the creation operators to the left of all annihilation operators. However, products of the exchange factors appear, for example
\begin{equation}\label{supp: normal-order}
\begin{split}
    &:a(\tilde p_1)a^\dagger(\tilde p_2):\, = \mathcal{Q}_\theta(p_1,p_2) a^\dagger(\tilde p_2) a(\tilde p_1), \\
    &:a(\tilde p_1)a^\dagger(\tilde p_2)a(\tilde p_3)a^\dagger(\tilde p_4):\, = \mathcal{Q}_\theta(p_1,p_2) \mathcal{Q}_\theta(p_3,p_4) \mathcal{Q}_\theta(p_1,p_4) a^\dagger(\tilde p_2)a^\dagger(\tilde p_4)a(\tilde p_1)a(\tilde p_3) ,
\end{split}
\end{equation}
and we have for the contraction of $a$ and $a^\dagger$ 
\begin{equation}
    \wick{ \c1 a(\tilde p_1) \c1 a^\dagger(\tilde p_2)} = a(\tilde p_1) a^\dagger(\tilde p_2)\, - :a(\tilde p_1) a^\dagger(\tilde p_2): \,= \delta(\tilde p_1,\tilde p_2) .
\end{equation}
Note that in this framework the original relative order of the creation operators among themselves must be preserved, and the same holds for the annihilation operators as well. For a term with no contractions, like \eqref{supp: normal-order}, the number of $\mathcal{Q}_\theta$ is the inversion number of the permutation from the original order to the normal-ordered form. For a term with contractions, the number of $\mathcal{Q}_\theta$ associated with a given contraction is the inversion number of the permutation which brings the contracted pair next to each other, for example
\begin{equation}
    \wick{ :\c1 a(\tilde p_1)a(\tilde p_2) \c1 a^\dagger(\tilde p_3)a^\dagger(\tilde p_4): \, = \mathcal{Q}_\theta(p_2,p_3) \c1 a(\tilde p_1) \c1 a^\dagger(\tilde p_3) \,:a(\tilde p_2) a^\dagger(\tilde p_4): },
\end{equation}
then the contracted pair is removed from the product and the procedure continues
\begin{equation}
    \wick{ :\c1 a(\tilde p_1)a(\tilde p_2) \c1 a^\dagger(\tilde p_3)a^\dagger(\tilde p_4): \, = \delta(\tilde p_1,\tilde p_3) \mathcal{Q}_\theta(p_2,p_3) \mathcal{Q}_\theta(p_2,p_4) a^\dagger(\tilde p_4)a(\tilde p_2)}.
\end{equation}
The number of $\mathcal{Q}_\theta$ associated with contraction of a given pair depends on the order in which the pairs are contracted 
\begin{equation}
\begin{split}
    &\wick{ :\c2 a(\tilde p_1) \c1 a(\tilde p_2) \c1 a^\dagger(\tilde p_3) \c2 a^\dagger(\tilde p_4):\propto 1} ,\\
    &\wick{ :\c1 a(\tilde p_1) \c2 a(\tilde p_2) \c1 a^\dagger(\tilde p_3) \c2 a^\dagger(\tilde p_4):\propto \mathcal{Q}_\theta(p_2,p_3)};    
\end{split}
\end{equation} 
however, their total number in the normal-ordered expansion does not depend on the order in which the contractions are performed.

Now, we focus on two-point functions. Let $\Phi$ be a $\mathcal{Q}_\theta$-deformed charged scalar field. The $\mathcal{Q}_\theta$-deformed commutator of its positive- and negative-frequency parts satisfies

\begin{equation}
    [\Phi^+(x),\Phi^{-\dagger}(y)]_{\mathcal{Q}_\theta} = [\Phi^{+\dagger}(x),\Phi^-(y)]_{\mathcal{Q}_\theta} = \Delta^+(x-y)  ,
\end{equation}

\noindent where 

\begin{equation}
	\Delta^+(x) = \frac{1}{(2\pi)^3}\int \frac{d^3p}{2E_p} e^{-ip \cdot x} .
\end{equation}
The free propagator 
\begin{equation}
\begin{split}
    \Delta(x-y) &= \braket{0|T\{\Phi(x)\Phi^\dagger(x')\}|0} \\
    &= \theta(t-t')\braket{0|\Phi(x)\Phi^\dagger(x')|0} + \theta(t'-t)\braket{0|\Phi^\dagger(x')\Phi(x)|0} ,
\end{split}
\end{equation}
where $\theta(t)$ is the Heaviside function, has the usual integral representation. This result extends to interacting fields and to fields with arbitrary spin.

\section{Relativistic wave equations for atoms}\label{supp: sec: atomic states}

QFT provides the appropriate relativistic framework to describe bound states, which appear as poles of Green functions. These functions describe the propagation of multiparticles systems and generalize the Dirac (or Breit) equation through the Bethe-Salpeter (BS) equation \cite{BetheSalpeter1951,salpeter1952mass,nakanishi1969general}, which encodes the full relativistic interaction among the particles forming the bound state. We employ this approach to analyze $\mathcal{Q}_\theta$-deformed atomic systems, focusing on the helium atom as the simplest configuration exhibiting violations of ordinary fermionic statistics.

Let us begin with the commutative framework. The relativistic definition of a wave function is given by the BS amplitude, which for the helium-like atom reads
\begin{equation}\label{supp: ordinary BS amplitude}
    \phi_{BS}(x_1,x_2,x_n;P) = \bra{\Omega}T\{ \Psi_e(x_1)\Psi_e(x_2)\Psi_n(x_n) \} \ket{\phi;P} ,
\end{equation}
where $\ket{\phi;P}$ denotes the bound state with total four-momentum $P^\mu$ (other quantum numbers are omitted). Here, $\Psi_{e,n}$ are the Heisenberg fields of the electron and nucleus, respectively, and $\ket{\Omega}$ is the vacuum state of the interacting theory. The corresponding six-point Green function is
\begin{equation}\label{supp: 6-point green function}
    G_0(x_1,x_2,x_n,x_1',x_2',x_n') = \braket{\Omega|T\{ \Psi_e(x_1)\Psi_e(x_2)\Psi_n(x_n)\Psi_e^\dagger(x_1')\Psi_e^\dagger(x_2')\Psi_n^\dagger(x_n') \}|\Omega}.
\end{equation}
The BS amplitude is antisymmetric under exchange of the electrons and the Green function shares this antisymmetry in both its initial and final electron coordinates. By inserting a complete set of states into Eq. \eqref{supp: 6-point green function} through 
\begin{equation}
\begin{split}
    \mathbb{I} &= \sum_n \ket{n}\bra{n} \\
    &= \sum_{\substack{\text{bound} \\ \text{states}}}\int \frac{d^4Q}{(2\pi)^4} \delta(P^2-M_B^2)\theta(P_0) \ket{\phi;Q}\bra{\phi;Q} + \text{“1-particle”} + \cdot\cdot\cdot , 
\end{split}
\end{equation}
and isolating the specific contribution of our bound state, one obtains
\begin{equation}\label{supp: green function 1}
\begin{split}
    G_0(\{x_i\},\{x'_i\}) = \int \frac{d^3Q}{(2\pi)^32E_Q} \Big[ \phi_{BS}(\{x_i\};Q)\phi_{BS}^{\dagger}(\{x_1'\};Q) \\
    \times\theta(\min(x_1^0,x_2^0,x_n^0) - \min(x_1'^0,x_2'^0,x_n'^0)) \Big]  ,
\end{split}
\end{equation}
where $\{x_i\} = x_1,x_2,x_n$, etc. In momentum space, the Green function has a pole at $P^0=E_P$, and in the limit $P^0\to E_P$ it becomes  
\begin{equation}\label{supp: Green function}
    G_0(\{p_i\},\{p_i'\}) = \frac{i\phi_{BS}(\{p_i\})\phi_{BS}^{\dagger}(\{p_i'\})}{2E_P[P^0-E_P+i\epsilon]}.
\end{equation}

In the interaction picture, the Green function can be written as
\begin{equation}
    G_0(\{x_i\},\{x_i'\}) = \frac{\bra{0}T\{\Psi_I^e(x_1)\Psi_I^e(x_2)\Psi_I^n(x_n)\Psi_I^{e\dagger}(x_1')\Psi_I^{e\dagger}(x_2')\Psi_I^{e\dagger}(x_n') e^{i\int d^4x \mathcal{H}_I}\}\ket{0}}{\bra{0}e^{i\int d^4x \mathcal{H}_I}\ket{0}} ,
\end{equation}
where $\mathcal{H}_I$ is the interaction Hamiltonian density. Expanding $G$ perturbatively yields a series of connected Feynman diagrams. The symmetry properties of the Green function under electron exchange are entirely determined by those of the free fields and are unaffected by the specific form of $\mathcal{H}_I$. The Green function describing the atom propagation can thus be decomposed as
\begin{align}
    &G_0(\{p_i\};\{p'_i\}) = S_{e}(p_1,p_2)S_n(p_n)\mathcal{I}_0 S_n(p_n')S_e(p_1',p_2')   ,\label{supp: Green function decomposition} \\
    &S_e(p_1,p_2) \coloneqq S_e(p_1)S_e(p_2) - S_e(p_2)S_e(p_1) ,
\end{align}
where $S_{e,n}$ denote the dressed propagators for the electrons and the nucleus, and $\mathcal{I}_0$ is the amputated correlation function. 

To proceed, let us express the amplitude in terms of an unsymmetrized amplitude $\hat\phi_{BS}$ obeying infinite statistics
\begin{equation}
    \phi_{BS}(p,q) = \hat\phi_{BS}(p,q) - \hat\phi_{BS}(q,p).
\end{equation}
Using Eqs. \eqref{supp: Green function} and \eqref{supp: Green function decomposition}, one finds
\begin{equation}\label{supp: no name}
    \frac{i\hat\phi_{BS}(\{p_i\})\hat\phi_{BS}^\dagger(\{p_i'\})}{2E_P[P^0-E_P+i\epsilon]} = S_e(p)S_e(q)S_n(p_n)\mathcal{I}_0S_e(p')S_e(q')S_n(p_n') .
\end{equation}
We stress that so far the explicit form of the interaction Hamiltonian has not been specified. 

\begin{figure}[t]
	\centering
	\includegraphics[width=0.8\linewidth]{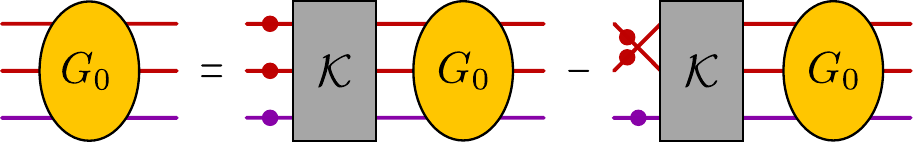}
	\caption{Structure of the amputated correlation function $\mathcal{I}$ for a helium-like atom. Red and purple lines denote dressed electron and nucleus propagators, respectively. The kernel $\mathcal{\mathcal{\mathcal{K}}}$ collects all irreducible two-body and three-body interactions.}
	\label{supp: fig: irreducible diagrams}
\end{figure}

By reorganizing diagrams, all two-body and three-body irreducible interactions can be isolated into a kernel $\mathcal{K}$, as shown in Figure \eqref{supp: fig: irreducible diagrams}, leading to
\begin{equation}\label{supp: no name 2}
\begin{split}
    \frac{i\hat\phi_{BS}(\{p_i\})\hat\phi_{BS}^\dagger(\{p_i'\})}{2E_P[P^0-E_P+i\epsilon]} = S_e(p_1)S_e(p_2)S_n(p_n) \int \prod_{i=1}^3d^4 k_i \,\mathcal{K}(\{p_i\};\{k_i\}) \\
    \times\left[S_e(k_1,k_2)S_n(k_n)\mathcal{I}_0 S_n(p_n')S_e(p_1')S_e(p_2')\right] .
\end{split}
\end{equation}
Substituting \eqref{supp: no name} into the square brackets of Eq. \eqref{supp: no name 2} and antisymmetrizing with respect to $p_1$ and $p_2$ yields the homogeneous BS equation
\begin{equation}\label{supp: 3p BS equation}
    S^{-1}_e(p_1)S^{-1}_e(p_2)S^{-1}_n(p_n)\phi_{BS}(\{p_i\}) = \int \prod_{i=1}^3 d^4 k_i \,\mathcal{K}'(\{p_i\};\{k_i\})\phi_{BS}(\{k_i\}) .
\end{equation}
This is the exact covariant equation for a helium-like system. 

Note that the amplitudes $\phi_{BS}$ and $\hat\phi_{BS}$ depend on the four-momenta $p_i^\mu = (E_i,\bb p_i)$, which satisfy $\sum_{i} p_i^\mu = P^\mu$. Owing to time-translation invariance, the amplitudes can be expressed in terms of relative energies (or, equivalently, relative times $x_i^0-x_j^0$ in configuration space). This feature represents a key difference from the Schr\"odinger or Dirac equations, which involve only a single time variable. To eliminate the dependence on relative energies, Eq. \eqref{supp: 3p BS equation} can be reduced to a three-dimensional equation—the Salpeter equation \cite{salpeter1952mass}. The multi-time amplitude $\phi_{BS}$ can be reconstructed from a single-time wave function $\psi(t,\{\bb p_i\})$ by propagating it to all possible relative time orderings. As shown in Ref. \cite{feldman1982relation}, this is achieved by applying one- and two-particle propagators to $\psi$ as follows
\begin{equation}\label{supp: multi to single time amplitude}
    \phi_{BS} = -\frac{1}{4\pi} \left[ (S_1+S_2)G_{12} + (S_2+S_3)G_{23} + (S_1+S_3)G_{13}  \right] \psi ,
\end{equation}
where $G_{ij}$ denotes the two-body Green function propagating particles $i$ and $j$. Neglecting retardation effects—justified by the dominance of instantaneous interactions—and retaining only two-particle interactions for simplicity, so that $\mathcal{K}(\{p_i\};\{p_i'\}) \to\mathcal{K}(\{\bb p_i\};\{\bb p_i'\}) \to \sum_{i>j}\mathcal{V}_{ij}(\{\bb p_i\};\{\bb p_i'\})$, one obtains the Salpeter equation \cite{mittleman1989three}
\begin{equation}\label{supp: salpeter eq}
\begin{split}
    &\left[\sum_{i=1}^3 h_D^{(i)} + \sum_{i>j}V_{ij}^+ - E \right]\psi=0 ,\\
    &(V^+_{ij} \psi)(\bb p, \bb q) = \int d^3kd^3l \, \mathcal{V}^+_{ij} (\bb p,\bb q;\bb k,\bb l)\psi(\bb k, \bb l) .
\end{split}
\end{equation}
Here $\psi(\bb p,\bb q) = \hat\psi(\bb p,\bb q) - \hat\psi(\bb q,\bb p)$ contains only positive-energy states, and $V_{ij}^+ \coloneqq \Lambda_+^{(i)}\Lambda_+^{(j)} V_{ij} \Lambda_+^{(i)}\Lambda_+^{(j)}$, where $\Lambda_+^i$ are projectors onto the positive energy spectrum of $h_D^{(i)} = \bb\alpha_{(i)}\cdot \bb p_i + \beta_{(i)} m_i$. Eq. \eqref{supp: salpeter eq} is the relativistic wave equation for helium-like atoms.

\subsubsection*{Noncommutative setting}

\begin{figure}[t]
	\centering
	\includegraphics[width=\linewidth]{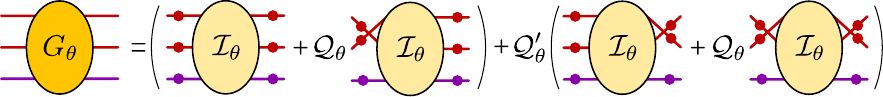}
	\caption{Six-point Green function describing the propagation of a helium-like atoms. The blobbed lines are the dressed propagators of the electrons (red) and the nucleus (purple), $\mathcal{I}_\theta$ is the amputated correlation function.}
	\label{supp: fig: Green function}
\end{figure}

Now we turn to the noncommutative generalization. In the general case where a quon deformation $\eta$ is present, a straightforward extension of the above derivation encounters difficulties. As illustrated in Figure \ref{supp: fig: Green function}, the external legs involve the exchange factor $\mathcal{Q}_\theta$ whenever electron permutations occur. Consequently, the expansion of the Green function modifies Eq. \eqref{supp: Green function decomposition} into
\begin{equation}
    S_e^{(\mathcal{Q}_\theta)}(p_1,p_2) \coloneqq S_e(p_1)S_e(p_2) + \mathcal{Q}_\theta(p_1,p_2) S_e(p_2)S_e(p_1) ,
\end{equation}
which has no definite symmetry. 

On the other hand, inserting a complete set of states into \eqref{supp: 6-point green function}, and proceeding as in Eq. \eqref{supp: green function 1}, yields BS amplitudes that inherit the symmetry of the bound state under consideration. Indeed, physical states may retain a (twisted) symmetry even including a quon deformation; the key novelty is that all irreducible representation of the symmetric group contribute (for helium, the twisted-symmetric and twisted-antisymmetric representations). Hence, the BS amplitude naturally reflects the symmetry of the physical state annihilated by the fields. As an example, consider a twisted-symmetric state of two free scalar particles. The equal-time BS amplitude is
\begin{equation}
\begin{split}
    \phi_{BS}^{free}(\bb{x},\bb{y},t) &= \braket{0|\Phi(\bb{x},t)\Phi(\bb{y},t)|\phi_{s_\theta}} \\
    &= \frac{1}{(2\pi)^3}\int \frac{d^3k}{\sqrt{2E_k}}\frac{d^3l}{\sqrt{2E_l}} e^{i(E_k+E_l)t}e^{-i \bb k \cdot \bb x}e^{-i \bb l\cdot \bb y} \left[\phi_{s_\theta}(\bb k,\bb l) + \mathcal{Q}_\theta(l,k) \phi_{s_\theta}(\bb l,\bb k)\right] \\
     &= \frac{1}{(2\pi)^3}\int \frac{d^3k}{\sqrt{2E_k}}\frac{d^3l}{\sqrt{2E_l}} e^{i(E_k+E_l)t}e^{-i \bb k \cdot \bb x}e^{-i \bb l\cdot \bb y}  \phi_{s_\theta}(\bb k,\bb l) \left[1 + \eta(l,k) \right] .
\end{split}
\end{equation}
In Fourier space this becomes
\begin{equation}\label{supp: free amplitude}
    \phi_{BS}^{free}(\bb p,\bb q, t) \propto e^{i(E_p+E_q)t} \phi_{s_\theta}(\bb p,\bb q) \left[1 + \eta(p,q) \right] = f_\theta^2(q,p) \phi_{BS}^{free}(\bb q,\bb p, t) .
\end{equation}
Therefore, each amplitude can be labeled according to the symmetry type $\gamma_\theta$ of its associated state
\begin{equation}
    \phi_{BS}^{\gamma_\theta}(x_1,x_2,x_n;P) = \braket{\Omega| T\{\Psi_e(x_1)\Psi_e(x_2)\Psi_n(x_n) \} |\phi;\gamma_\theta,P}.
\end{equation}

To derive an equation for the BS amplitude analogous to \eqref{supp: 3p BS equation}, we then consider a suitable linear combination of Green functions
\begin{equation}\label{supp: super green function}
\begin{split}
     \mathcal{G}_{\gamma_\theta}(\{p_i\};\{p_i'\}) = \frac{1}{4}\Big[ &G_\theta(p_1,p_2,p_n;p_1',p_2',p_n') + s_{\gamma_\theta} f^2_\theta(p_2,p_1) G_\theta(p_2,p_1,p_n;p_1',p_2',p_n') \\
     s_{\gamma_\theta}f^2_\theta(p_2',p_1')\Big(&G_\theta(p_1,p_2,p_n;p_2',p_1',p_n') + s_{\gamma_\theta} f^2_\theta(p_2,p_1) G_\theta(p_2,p_1,p_n;p_2',p_1',p_n')\Big)\Big] ,
\end{split}
\end{equation}
where $s_{\gamma_\theta}$  denotes the sign of the representation: $+1$ for the symmetric and $-1$ for the antisymmetric case. Using Eq. \eqref{supp: Green function} to expand each component in terms of BS amplitudes gives
\begin{equation}
    \mathcal{G}_{\gamma_\theta}(\{p_i\};\{p_i'\}) = \left(\frac{1+s_{\gamma_\theta}\eta(p_1,p_2)}{2}\right)\left(\frac{1+s_{\gamma_\theta}\eta(p_1',p_2')}{2}\right) \frac{i\phi_{BS}^{\gamma_\theta}(\{p_i\})\phi_{BS}^{\gamma_\theta\dagger}(\{p_i'\})}{2E_P[P^0-E_P+i\epsilon]} ,
\end{equation}
where $\phi_{BS}^{\gamma_\theta}(\{p_i\}) = \hat\phi_{BS}(p_1,p_2,p_n) + s_{\gamma_\theta}f_\theta^2(p_2,p_1) \hat\phi_{BS}(p_2,p_1,p_n)$. On the other hand, using  
\begin{equation}
    G_\theta(\{p_i\};\{p'_i\}) = S_{e}^{(\mathcal{Q}_\theta)}(p_1,p_2)S_n(p_n)\mathcal{I}_\theta S_n(p_n')S_e^{(\mathcal{Q}_\theta)}(p_1',p_2')   ,
\end{equation}
we obtain
\begin{equation}
\begin{split}
    &\mathcal{G}_{\gamma_\theta}(\{p_i\};\{p_i'\}) = \left(\frac{1+s_{\gamma_\theta}\eta(p_1,p_2)}{2}\right)\left(\frac{1+s_{\gamma_\theta}\eta(p_1',p_2')}{2}\right)  \\ 
    &\qquad\qquad\qquad\qquad \times S_e^{(\gamma_\theta)}(p_1,p_2)S_n(p_n)\mathcal{I}_\theta S_n(p_n') S_e^{(\gamma_\theta)}(p_1',p_2') ,
\end{split}
\end{equation}
with
\begin{equation}
    S_e^{(\gamma_\theta)}(p_1,p_2) \coloneqq S_e(p_1)S_e(p_2) + s_{\gamma_\theta}f^2_\theta(p_2,p_1) S_e(p_2)S_e(p_1) ,
\end{equation}
which has the appropriate symmetry. We finally obtain
\begin{equation}
    \frac{i\phi_{BS}^{\gamma_\theta}(\{p_i\})\phi_{BS}^{\gamma_\theta\dagger}(\{p_i'\})}{2E_P[P^0-E_P+i\epsilon]}  = S_e^{(\gamma_\theta)}(p_1,p_2)S_n(p_n) \mathcal{I}_\theta S_n(p_n) S_e^{(\gamma_\theta)}(p_1',p_2') .
\end{equation}

It is now tempting to write the BS amplitude in terms of the function $\hat\phi_{BS}$ and derive an equation analogous to \eqref{supp: 3p BS equation}. However, this requires some care. Once the kernel is isolated within the amputated correlation function $\mathcal{I}_\theta$, we wish to construct an interaction such that the amplitude enters with the appropriate symmetry $\gamma_\theta$. To verify whether this is the case, one would need explicit knowledge of the interaction Hamiltonian $\mathcal{H}_I^\theta$, which, in the presence of quon deformations, is not known. 

This observation leaves two distinct possibilities: either the electron-electron interaction part of the kernel preserves statistics, or it does not. In the first case, after reducing the BS amplitude to a single-time amplitude, one obtains the Salpeter equation
\begin{equation}\label{supp: Salpeter 1}
    \left[\sum_{i=1}^3 h_D^{(i)} + \sum_{i>j}V_{\theta,ij}^+ - E \right]\psi_{\gamma_\theta}=0 ,
\end{equation}
where again $V^+_{\theta,ij}$ are the electron-electron and electron-nucleus interactions, now modified by noncommutative corrections. Physical states then define a subset of the irreducible subspaces $\mathcal{H}_{\gamma_\theta}^2$. In the second case, where the electron-electron interaction breaks permutation symmetry, a closed equation can be obtained only for the non-symmetrized amplitude. The corresponding single-time amplitude then satisfies
\begin{equation}\label{supp: Salpeter 2}
    \left[\sum_{i=1}^3 h_D^{(i)} + \sum_{i>j}V_{\theta,ij}'^{+} - E \right]\psi_{\mathcal{Q}_\theta} =0 ,
\end{equation}
where $\psi_{\mathcal{Q}_\theta}(\{\bb p_i\}) = \psi(\bb p_1,\bb p_2,\bb p_n) + \mathcal{Q}_\theta(p_1,p_2)\psi(\bb p_2,\bb p_1,\bb p_n)$. 

Both Eqs. \eqref{supp: Salpeter 1} and \eqref{supp: Salpeter 2} reduce to the standard QED expression in the limit $\Lambda_\theta\to\infty$. Consequently, to lowest order in $\Lambda_\theta^{-1}$, the eigenfunctions coincide with the ordinary ones. These equations, as well as their $n$-electron generalization, allow to compute noncommutative corrections to atomic spectra. However, for the purposes of our analysis, the explicit expressions of Eqs. \eqref{supp: Salpeter 1} and \eqref{supp: Salpeter 2} are not required.

\section{Evaluation of the rate without SSRs}\label{supp: sec: rates}

In the following, we derive the expressions for the PEP-violating rates without assuming SSRs (Eq.  \eqref{rate}). We recall the expression for the PEP-violating transition matrix elements, which have two contributions $H_{if}^\pm = H_{if}^a \pm H_{if}^s$ given by
\begin{equation}\label{supp: no SSRs matrix element}
\begin{split}
    H_{if}^\gamma = \frac{e}{\sqrt{2\omega_k}} \int d^3pd^3q D_\theta^\gamma\psi_{s}'^\dagger(\bb p-\bb k,\bb q)  \bb\alpha_{(1)}\cdot\bb\varepsilon^*_\lambda \psi_\gamma(\bb p,\bb q) ,
\end{split}
\end{equation}
with symmetric and antisymmetric deformations
\begin{align}\label{supp: deformation no SSRs} 
    &D_\theta^s(p,q,k) = \frac{[1+\mathcal{Q}_\theta(p,q)][1+\mathcal{Q}_\theta^*(p-k,q)] + 1 - \eta^2(p-k,q)}{\Vert{\psi'_s}\Vert\Vert{\psi}\Vert} , \\
    &D_\theta^a(p,q,k) = \frac{[1-\mathcal{Q}_\theta(p,q)][1+\mathcal{Q}_\theta^*(p-k,q)] - 1 + \eta^2(p-k,q)}{\Vert{\psi'_s}\Vert\Vert{\psi}\Vert} .
\end{align}
In addition, the nonrelativistic expression for the function $\eta$ is 
\begin{equation}\label{supp: eta expansion}
    \eta(p,q) \sim \pm \left[1 - \left(\frac{m^{1-a} |\bb p - \bb q|^a}{\Lambda_\theta}\right)^\kappa \right],
\end{equation}
where $\kappa>0$ and $a\in\mathbb{R}$.

The $+$ ($-$) sign in $H_{if}^\pm$ corresponds to the amplitude $\psi(\tilde{p},\tilde{q})$ ($\psi(\tilde{q},\tilde{p})$). In general, permuting the electrons leads to different expressions for the squared matrix element $|H_{if}^\pm|^2 = |H_{if}^a \pm H_{if}^s|^2$. Only when a single component $H_{if}^\gamma$ dominates one can approximate $|H_{if}^\pm|^2\approx |H_{if}^\gamma|^2$, in which case the result becomes effectively independent of the electrons permutation.

In order to compute the matrix elements \eqref{supp: no SSRs matrix element}, we begin by giving the expression of the PEP-violating state norm 
\begin{equation}\label{supp: PEP-violating norm exact}
\norm{\psi_s'}^2 = \int_{p,q} \left(\frac{1-\eta(p,q)\cos(p \wedge_{\theta} q)}{2}\right) |\psi_s'(\bb p,\bb q)|^2 ,
\end{equation}
where $\int_p = \int d^3p$. To lowest order in $\Lambda_\theta^{-1}$, the norm $\norm{\psi_s'}$ depends on the expansion of $\eta$ given in Eq. \eqref{supp: eta expansion}, which yields
\begin{align}
    \eta=-1 \text{ or } \kappa>4 :&\quad \norm{\psi_s'}^2\sim \frac{1}{\Lambda_\theta^4} \int_{p,q}  |\psi_s'^{(0)}(\bb p,\bb q)|^2  \frac{(p \wedge_c q)^2}{4} ,\label{supp: PEP-violating norm leading 1} \\[0.2cm]
     0<\kappa<4: &\quad \norm{\psi_s'}^2 \sim \frac{1}{\Lambda_\theta^\kappa} \int_{p,q}|\psi_s'^{(0)}(\bb p,\bb q)|^2 \frac{\sigma^\kappa(p,q)}{2} , \label{supp: PEP-violating norm leading 2}\\[0.2cm]
     \kappa=4: &\quad \norm{\psi_s'}^2 \sim \frac{1}{\Lambda_\theta^4} \int_{p,q}|\psi_s'^{(0)}(\bb p,\bb q)|^2 \left[\frac{2\sigma^4(p,q)+(p \wedge_c q)^2}{4}\right]  ,\label{supp: PEP-violating norm leading 3}
\end{align}
where $p \wedge_c q \coloneqq p_\mu c^{\mu\nu} q_\nu$, where we recall that $\theta^{\mu\nu}=c^{\mu\nu}/\Lambda_\theta^2$. For convenience, we define $\mathcal{N}_\kappa$ as the $\Lambda_\theta$-independent integrals appearing above, such that $\norm{\psi_s} = \Lambda_\theta^{-\frac{\kappa}{2}}\mathcal{N}_\kappa$ for $0<\kappa<4$ and $\norm{\psi_s} = \Lambda_\theta^{-2}\mathcal{N}_\kappa$ for $\eta=-1$ or $\kappa\geq4$. 

In general, a $n$-electron wave function $\psi^{(0)}(\bb p_1,...,\bb p_n)$ can be expanded as a linear combination of hydrogen-like wave functions $\varphi_\alpha$
\begin{equation}
    \psi^{(0)}(\bb p_1,...,\bb p_n) = \sum_{\alpha_1,...,\alpha_n} A_{\alpha_1,...,\alpha_n} \varphi_{\alpha_1}(\bb p_1)\cdot\cdot\cdot\varphi_{\alpha_n}(\bb p_n),
\end{equation}
where each index $\alpha$ represents both discrete and continuous quantum numbers; in the latter case, the summation symbol $\sum_{\alpha_1,...,\alpha_n}$ is understood as an integration over the continuous indices. In general, no symmetrization is imposed on the coefficients $A_{\alpha_1,...,\alpha_n}$, since, if SSRs are violated, physical states do not possess definite permutation symmetry. 

For our purposes, it is sufficient to approximate the wave functions by retaining only their dominant contribution, say 
\begin{equation}\label{supp: wave function dominant contribution}
    \psi^{(0)}(\bb p_1,...,\bb p_n) = \varphi_{\alpha_1}(\bb p_1)\cdot\cdot\cdot\varphi_{\alpha_n}(\bb p_n) .
\end{equation}
In the non-relativistic limit, contributions of the four-spinors $\varphi_\alpha = (\phi_\alpha\ \chi_\alpha)^T$ come only from the upper component
\begin{equation}
    \phi_\alpha(\bb p) = \frac{1}{p_0^{3/2}} R_\alpha(|\bb p|)\Omega_\alpha(\hat{\bb p}),
\end{equation}
since $\chi_\alpha\sim \frac{p}{m}\phi_\alpha$ is suppressed. As an example 
\begin{equation}
    \varphi_\beta^\dagger(\bb p) \bb\alpha\cdot\bb\varepsilon^*_\lambda \varphi_\alpha(\bb p+\bb k) \approx \frac{\bb\varepsilon^*_\lambda\cdot\bb p}{m}\phi_\beta^\dagger(\bb p)\phi_\alpha(\bb p+\bb k) .
\end{equation}
The radial and angular parts of the upper component $\phi_\alpha$ are given by \cite{podolsky1929momentum}
\begin{align}
    &R_{nl}(|\bb p|) = \sqrt{\frac{2}{\pi}\frac{(n-l-1)!}{(n+1)!}} n^2 2^{2l+2} l! \frac{n^l (p/p_0)^l}{(n^2(p/p_0)^2+1)^{l+2}} \mathcal{C}^{(l+1)}_{n-l-1}\left(\frac{n^2(p/p_0)^2-1}{n^2(p/p_0)^2+1}\right) , \label{supp: radial part}\\
    &\Omega_{\kappa j m}(\hat{\bb p}) = \sum_s C(l\frac{1}{2}j;m-s,s)Y_{l, m-s}(\hat{\bb p})\xi_s  ,\label{supp: angular part}
\end{align}
where $\mathcal{C}^{(l+1)}_N$ are Gegenbauer polynomials, $C(l\frac{1}{2}j;m-s,s)$ Clebsh-Gordan coefficients, $Y_{l, m-s}$ spherical harmonics and $\xi_s$ spin eigenfunctions. $p_0 = Z\hbar /a_0$ is the momentum of the first Bohr orbit and $a_0$ the Bohr radius. For later convenience, we have extracted from the function $R_{nl}$ the normalization factor $p_0^{-3/2}$.

\subsection{Unsuppressed scenarios: $\kappa \geq 4$}

The leading contribution to the transition matrix element \eqref{supp: no SSRs matrix element} arises from the contribution $H_{if}^a$ involving the antisymmetric component $\psi_a$ of the initial state. This follows because the numerator of $D^s_\theta$ scales as $\mathcal{O}(\Lambda_\theta^{-4})$ and the numerator of $D^a_\theta$ goes like $\mathcal{O}(\Lambda_\theta^{-2})$. From Eq. \eqref{supp: PEP-violating norm leading 1}, the norm of the final PEP-violating state scales as $\mathcal{O}(\Lambda_\theta^{-2})$, and therefore the deformation term $D^a_\theta$ given in Eq. \eqref{supp: deformation no SSRs} becomes independent of the noncommutativity scale
\begin{equation}
    D_\theta^a \sim \frac{(p-k) \wedge_c q}{\mathcal{N}_{\kappa\geq4}} \sim \mathcal{O}(1) .
\end{equation}
The resulting matrix element then reads 
\begin{equation}\label{supp: matrix element kappa=0}
     H_{if}^\pm = \pm\frac{2ie}{\sqrt{2\omega_k}\mathcal{N}_{\kappa\geq4}}\bb\varepsilon_\lambda^*\cdot \int_{p,q} \psi'^{(0)\dagger}_s(\bb p - \bb k,\bb q) \bb\alpha_{(1)} \psi_a^{(0)}(\bb p,\bb q)\, (p-k) \wedge_c q .
\end{equation}
Crucially, this matrix element—and hence the corresponding transition rate—is independent of the noncommutative scale $\Lambda_\theta$. As we shall discuss below, these matrix elements do not vanish in general and yield rates proportional to the standard rates.

\subsection{Maximal suppression: $\kappa=2$}

Now, we dedicate particular care to the scenarios that yield the maximal suppression of the PEP-violating rates, corresponding to a quon deformation $\eta+1$ quadratically suppressed with the noncommutativity scale. In this case, both the components $H_{if}^s$ and $H_{if}^a$ contribute at leading order, yielding the matrix elements
\begin{equation}\label{supp: matrix element kappa=2}
     H_{if}^\pm =  \frac{2}{\mathcal{N}_2\Lambda_{\theta}}\frac{e}{\sqrt{2\omega_k}}\bb\varepsilon_\lambda^*\cdot \int_{p,q} \psi'^{(0)\dagger}_s(\bb p-\bb k,\bb q) \bb\alpha_{(1)}\left[ \psi_s^{(0)}(\bb p,\bb q) \sigma^2(p-k,q) \pm i \psi_a^{(0)}(\bb p,\bb q) (p-k)\wedge_c q\right]  .
\end{equation}
The dependence on the noncommutativity scale is given by the factor $1/\Lambda_\theta$, such that the associated rates scale as $\mathcal{O}(\Lambda_\theta^{-2})$. Expanding the wave functions in hydrogen-like basis as in Eq. \eqref{supp: wave function dominant contribution}, yields $\psi^{(0)}(\bb p,\bb q) = \varphi_\alpha(\bb p)\varphi_\beta(\bb q)$ and $\psi'^{(0)}_s(\bb p,\bb q) = \varphi_\alpha(\bb p)\varphi_\alpha(\bb q)$, corresponding to a transition $\beta\to \alpha$. 

Therefore, in the nonrelativistic limit Eq. \eqref{supp: matrix element kappa=2} becomes
\begin{equation}\label{supp: NR matrix element kappa=2}
    H_{if}^\pm =  \frac{1}{\mathcal{N}_2\Lambda_{\theta}}\frac{e}{\sqrt{2\omega_k}m}\bb\varepsilon_\lambda^*\cdot \int_{p,q} \bb p \,\phi^{\dagger}_\alpha(\bb p-\bb k)\phi^{\dagger}_\alpha(\bb q) \left[ \phi_{[\beta}(\bb p)\phi_{\alpha]}(\bb q) \sigma^2(p-k,q) \pm i\phi_{(\beta}(\bb p)\phi_{\alpha)}(\bb q) (p-k)\wedge_c q\right]  ,
\end{equation}
where $\phi_{[\beta}\phi_{\alpha]} \coloneqq \phi_{\beta}\phi_{\alpha}+\phi_{\alpha}\phi_{\beta}$ yields the component $H_{if}^s$ and $\phi_{(\beta}\phi_{\alpha)} \coloneqq \phi_{\beta}\phi_{\alpha}-\phi_{\alpha}\phi_{\beta}$ yields the component $H_{if}^a$. The PEP-violating state norm \eqref{supp: PEP-violating norm leading 2} writes as
\begin{equation}\label{supp: NR norm kappa=2}
    \mathcal{N}_2 = \left[ \int_{p,q} |\phi_\alpha(\bb p)|^2|\phi_\alpha(\bb q)|^2 \sigma^2(p,q) \right]^\frac{1}{2} .
\end{equation}

To compute matrix elements, we define the auxiliary integrals 
\begin{align}
    & \tilde{\mathcal{R}}_{\alpha_1\alpha_2}^{(b)} \coloneqq \int d\left(\frac{p}{p_0}\right)\,\left(\frac{p}{p_0}\right)^{2+b} R_{n_1l_1}\left(\frac{p}{p_0}\right) R_{n_2l_2}\left(\frac{p}{p_0}\right) \\
    &\mathcal{R}_{\alpha_1\alpha_2} \coloneqq  \int d\left(\frac{p}{p_0}\right) \,\left(\frac{p}{p_0}\right)^2 p R_{n_1l_1}\left(\frac{p}{p_0}\right) R_{n_2l_2}\left(\frac{p}{p_0}\right) = \sqrt{2mE_0}\tilde{\mathcal{R}}^{(1)} ,\label{supp: radial integral}\\
    &\Omega_{\alpha_1\alpha_2}^i \coloneqq \int d\varphi d\theta\sin\theta \, \hat{p}^i\, \Omega_{\kappa_1j_1m_1}^\dagger(\hat{\bb p})\Omega_{\kappa_2j_2m_2}(\hat{\bb p}), \\
    &\Omega^{ij}_{\alpha_1\alpha_2} \coloneqq \int d\varphi d\theta\sin\theta \, \hat{p}^i \,\hat{p}^j\, \Omega_{\kappa_1j_1m_1}^\dagger(\hat{\bb p})\Omega_{\kappa_2j_2m_2}(\hat{\bb p}).
\end{align}
Numerical values for $\tilde{\mathcal{R}}^{(1)}_{\alpha\beta}$ are listed in Table \ref{supp: tab: numerical values}. Selection rules arise from angular integrals: $\Omega_{\alpha\beta}^i\neq0$ only for states with opposite parity, $\Omega^{ij}_{\alpha\beta}\neq0$ only for states with same parity, and $\Delta m_l = -1,0,1$ for the orbital angular momentum projection.

Now we can calculate the integrals involved in Eqs. \eqref{supp: NR matrix element kappa=2} and \eqref{supp: NR norm kappa=2}. Using the dipole approximation $\phi_\beta(\bb p +\bb k) \approx \phi_\beta(\bb p)$, the relevant integrals reduce to
\begin{align}
    &\int d^3p \bb\varepsilon^*_\lambda\cdot\bb p \phi_{\alpha_1}^\dagger(\bb p) \phi_{\alpha_2}(\bb p) = \bb\varepsilon^*_\lambda \cdot \bb\Omega_{\alpha_1\alpha_2}\mathcal{R}_{\alpha_1\alpha_2} , \label{supp: I1}\\
    &\int d^3p \bb\varepsilon^*_\lambda\cdot\bb p \phi_{\alpha_1}^\dagger(\bb p) \phi_{\alpha_2}(\bb p) p^i  = \sqrt{2mE_0} X_{\alpha_1\alpha_2}  \left( \bb\varepsilon^*_\lambda\cdot \bb\Omega^i_{\alpha_1\alpha_2} \mathcal{R}_{\alpha_1\alpha_2} \right) , \label{supp: I2}\\
    &\int d^3p \bb\varepsilon^*_\lambda\cdot\bb p \phi_{\alpha_1}^\dagger(\bb p) \phi_{\alpha_2}(\bb p) |\bb p|^2 = 2mE_0 X'_{\alpha_1\alpha_2}  \left( \bb\varepsilon^*_\lambda\cdot \bb\Omega_{\alpha_1\alpha_2} \mathcal{R}_{\alpha_1\alpha_2} \right) , \label{supp: I3} \\
    &\int d^3p \phi_{\alpha_1}^\dagger(\bb p) \phi_{\alpha_2}(\bb p) |\bb p |^2 = 2mE_0\tilde{\mathcal{R}}_{\alpha_1\alpha_2}^{(2)} \delta_{j_1j_2} \delta_{m_{l1}m_{l2}}\delta_{l_1l_2} , \label{supp: I4}
\end{align}
with
\begin{equation}
    X_{\alpha_1\alpha_2} \coloneqq \frac{\tilde{\mathcal{R}}_{\alpha_1\alpha_2}^{(2)}}{\tilde{\mathcal{R}}_{\alpha_1\alpha_2}^{(1)}} ,\qquad X'_{\alpha_1\alpha_2} \coloneqq \frac{\tilde{\mathcal{R}}_{\alpha_1\alpha_2}^{(3)}}{\tilde{\mathcal{R}}_{\alpha_1\alpha_2}^{(1)}} .
\end{equation}
Table \ref{supp: tab: numerical values} gives a list of numerical values for these coefficients.  
\begin{table}
    \centering
    \caption{Numerical values for $\tilde R_{\alpha_1\alpha_2}^{(1)}$, $X_{\alpha_1\alpha_2}$ and $X_{\alpha_1\alpha_2}'$ for selected $\alpha_1,\alpha_2$.}
    \begin{tabular}{|c|ccccc|}
    \hline\hline
         $(n_1l_1),(n_2l_2)$&  $1s,2p$&  $2p,3d$&  $1s,3p$&  $2s,3p$& $2s,4f$\\
         $\tilde{\mathcal{R}}_{\alpha_1\alpha_2}^{(1)}$&  $0.48$&  $0.33$&  $0.23$&  $0.21$& $0.14$\\
     \hline \hline 
         $(n_1l_1),(n_2l_2)$&  $1s,1s$&  $2p,2p$&  $1s,3d$&  $3d,3d$& $2p,4f$\\
         $X_{\alpha_1\alpha_2}$&  $1.18$&  $0.55$&  $0.53$&  $0.36$& $0.35$\\ \hline \hline
         $(n_1l_1),(n_2l_2)$&  $1s,2p$&  $1s,3p$&  $1s,4p$&  $2p,3d$& \\
         $X'_{\alpha_1\alpha_2}$&  $0.75$&  $1.0$&  $1.09$&  $0.22$& \\
     \hline \hline
    \end{tabular}
    \label{supp: tab: numerical values}
\end{table}

In particular, we focus on transitions $\ket{\phi_\beta} \to \ket{\phi_\alpha}$ between states with opposite parity, for example 2p $\to$ 1s. The contribution of the symmetric component of the initial state involves the functions $\sigma_\text{nr}(p,q) = m^{1-a}|\bb p - \bb q|^a$.  We perform explicit calculations for the cases $a=0,1/2,1$, which correspond to $\sigma_\text{nr}(p,q)= m,\ \sqrt{m|\bb p - \bb q|}$ and $|\bb p - \bb q|$. Roughly speaking, each term $|\bb p - \bb q|$ can be approximated by the Bohr momentum $p_0 =  \sqrt{2mE_0}$, where $E_0=Z^2 \, 13.6$ eV for an atom with atomic number $Z$. Finally, we evaluate the contribution of $H_{if}^a$, proportional to $p \wedge_c q $.

\subsubsection{Case $\sigma_\text{nr}=m$}

The matrix element takes the form $H_{if}^s =  \frac{m}{\Lambda_{\theta}}(H_{if})_0$, where $(H_{if})_0$ corresponds to the PEP-allowed transition $\beta \to \alpha$. Forgetting the contribution of $H_{if}^a$, the PEP-forbidden rates are then given by
\begin{equation}\label{supp: rate Sigma_1}
    \frac{d\Gamma_\text{PV}^s}{d\Omega} =  \left(\frac{m}{\Lambda_{\theta}}\right)^2\frac{d\Gamma_0}{d\Omega} ,
\end{equation}
where $\frac{d\Gamma_0}{d\Omega}$ is the rate for the associated PEP-allowed transition.

\subsubsection{Case $\sigma_\text{nr}=|\bb p - \bb q|$}

Using the expressions given in Eqs. \eqref{supp: I1},\eqref{supp: I2}, \eqref{supp: I3} and \eqref{supp: I4}, the component $H_{if}^s$ of the matrix elements \eqref{supp: NR matrix element kappa=2} can be written as 
\begin{equation}\label{supp: NR matrix element Sigma_2}
    H_{if}^s = \frac{2mE_0 }{\mathcal{N}_2\Lambda_{\theta}} \frac{e}{\sqrt{2\omega_k}m} \left[ X'_{\alpha\beta} \bb\varepsilon_\lambda^*\cdot\bb\Omega_{\alpha\beta} \mathcal{R}_{\alpha\beta}  +  X_{\alpha\alpha}\tilde{\mathcal{R}}_{\alpha\alpha} \sum_i \varepsilon_\lambda^{i*} \Omega_{\alpha\beta}^i( 1-2 \Omega_{\alpha\alpha}^{ii})\mathcal{R}_{\alpha\beta}  \right]  .
\end{equation} 
For $\alpha$ corresponding to an s state, $\Omega_{\alpha\alpha}^{xx}=\Omega_{\alpha\alpha}^{yy}=\Omega_{\alpha\alpha}^{zz}=\frac{1}{3}$ and hence $\sum_i \varepsilon_\lambda^{i*} \Omega_{\alpha\beta}^i( 1-2 \Omega_{\alpha\alpha}^{ii})=\frac{1}{3} \bb\varepsilon_\lambda^*\cdot\bb\Omega_{\alpha\beta}$. Substituting into Eq. \eqref{supp: NR matrix element Sigma_2} gives
\begin{equation}\label{supp: NR matrix element Sigma_2 bis}
    H_{if}^s = \frac{2mE_0 (X'_{\alpha\beta}  +  \frac{1}{3} \tilde{\mathcal{R}}_{\alpha\alpha}^{(2)} )}{\mathcal{N}_2\Lambda_{\theta}}  (H_{if})_0  .
\end{equation} 
In particular, $\tilde{\mathcal{R}}_{\alpha\alpha}^{(2)}=1$ when $\alpha$ corresponds to the 1s level. The norm of the PEP-violating state is
\begin{equation}
    \mathcal{N}_2 = \left[ 2\int_p|\phi_\alpha(\bb p)|^2 p^2 \right]^\frac{1}{2} = \sqrt{4mE_0  \tilde{\mathcal{R}}^{(2)}_{\alpha\alpha}}.
\end{equation}
We define the coefficients
\begin{equation}
    \mathcal{X}_{\alpha\beta} = \frac{\left(X'_{\alpha\beta}  +  \frac{1}{3} \tilde{\mathcal{R}}_{\alpha\alpha}^{(2)}\right)^2}{\tilde{\mathcal{R}}_{\alpha\alpha}^{(2)}} ,
\end{equation}
so that the resulting rates become
\begin{equation}\label{supp: rate Sigma_2}
    \frac{d\Gamma_\text{PV}^s}{d\Omega} =  \left(\frac{\mathcal{X}_{\alpha\beta} \,mE_0}{\Lambda_{\theta}^2}\right)\frac{d\Gamma_0}{d\Omega} .
\end{equation}

\subsubsection{Case $\sigma_\text{nr}=\sqrt{m|\bb p - \bb q|}$} 

The matrix elements $H_{if}^s$ take the form
\begin{equation}\label{supp: NR matrix element Sigma_3}
\begin{split}
     H_{if}^s &=  \frac{m}{\mathcal{N}_2\Lambda_{\theta}}\frac{e}{\sqrt{2\omega_k}m}\bb\varepsilon_\lambda^*\cdot \int_{p,q} \bb p  \sqrt{p^2+q^2-2 \bb p \cdot \bb q} \,\phi^{\dagger}_\alpha(\bb p)\phi^{\dagger}_\alpha(\bb q)\phi_{[\beta}(\bb p)\phi_{\alpha]}(\bb q) \\
     &=  \frac{m}{\mathcal{N}_2\Lambda_{\theta}}\frac{e}{\sqrt{2\omega_k}m}\bb\varepsilon_\lambda^*\cdot \int_{p,q} (\bb p + \bb q)  \sqrt{p^2+q^2-2 \bb p \cdot \bb q} \,\phi^{\dagger}_\alpha(\bb p)\phi_{\beta}(\bb p)\left|\phi_{\alpha}(\bb q)\right|^2 .
\end{split}
\end{equation}
Using the identity
\begin{equation}
    \frac{1}{3}\bb\nabla_q \left(p^2+q^2-2 \bb p \cdot \bb q \right)^{3/2} = (\bb q - \bb p) \sqrt{p^2+q^2-2 \bb p \cdot \bb q} ,
\end{equation}
one can see that Eq. \eqref{supp: NR matrix element Sigma_3} can be recast into the form
\begin{equation}
\begin{split}
    \delta H_{if}^s =  \frac{m}{\mathcal{N}_2\Lambda_{\theta}}\frac{e}{\sqrt{2\omega_k}m}\bb\varepsilon_\lambda^*\cdot \int_{p,q} \phi^{\dagger}_\alpha(\bb p)\phi_{\beta}(\bb p)\left|\phi_{\alpha}(\bb q)\right|^2 \Bigg[2 \bb p \sqrt{p^2+q^2-2 \bb p \cdot \bb q}  \\
    +  \frac{1}{3}\bb\nabla_q \left(p^2+q^2-2 \bb p \cdot \bb q \right)^{3/2} \Bigg]   .
\end{split}
\end{equation}
We consider transitions to the $\alpha=$ 1s level, since spherical symmetry simplifies the computation. The second integral vanishes due to the orthogonality of the angular components $\Omega_{\alpha}$ and $\Omega_\beta$ of the wave functions (Eq. \eqref{supp: angular part}): performing integration in $\bb p$ yields
\begin{equation}
    \frac{1}{3}\int dq q^2 R_{\alpha}(q)R_\beta(q)\frac{\partial K(q)}{\partial q} \bb e_q\int d\varphi d\cos\theta \Omega^\dagger_{\alpha}(\theta,\varphi) \Omega_\beta(\theta,\varphi) = 0
\end{equation}
where $K$ depends only on $q=|\bb q|$ and $\bb e_q$ denotes the radial unit vector.

Defining the coefficients
\begin{equation}
    Y_{1s,\beta} = \frac{\int dp p^3 R_{1s}(p)R_{\beta}(p) I(p)}{\mathcal{R}_{1s,\beta}} = 
    \begin{cases}
        1.16 \quad \beta=2p\\
        1.27 \quad \beta=3p
    \end{cases} 
\end{equation}
where the function $I(p,q)$ is given by
\begin{equation}
\begin{split}
    I(p) &= \int_q \left| \phi_{1s}(q) \right|^2  \sqrt{p^2+q^2-2pq\cos \theta} \\
    &= \frac{1}{2}\int dqq^2 [R_{1s}(q)]^2 \frac{(p+q)^3-|p-q|^3}{3pq}  ,
\end{split}
\end{equation}
the remaining integral yields
\begin{equation}
    \int_{p,q} \,\bb p \sqrt{p^2+q^2-2 \bb p \cdot \bb q} \,\phi^{\dagger}_\alpha(\bb p)\phi_\beta(\bb p) \left|\phi_\alpha(\bb q)\right|^2 =  Y_{1s,\beta}\bb\Omega_{1s,\beta} \mathcal{R}_{1s,\beta} .
\end{equation}
We thus find that Eq. \eqref{supp: NR matrix element Sigma_3} is proportional to the ordinary transition matrix element. 

The norm of the final state is
\begin{equation}
\begin{split}
    \mathcal{N}_2 &= \left( m\sqrt{2mE_0} \int_{p,q} R_{1s}^2(p) R_{1s}^2(q) \sqrt{p^2+q^2-2\bb p \cdot \bb q} \right)^\frac{1}{2} \\
    &= \left(\frac{35}{9\pi} m\sqrt{2mE_0}\right)^{\frac{1}{2}} .
\end{split}
\end{equation}  
Finally, we obtain the following expression for the PEP-violating rates
\begin{equation}\label{supp: rate Sigma_3}
    \frac{d\Gamma_\text{PV}^s}{d\Omega} = \frac{4\mathcal{Y}_{\alpha\beta}\sqrt{2m^3E_0}}{\Lambda_\theta^2}\frac{d\Gamma_0}{d\Omega} ,
\end{equation}
where $\mathcal{Y}_{\alpha\beta}$ combines $Y_{\alpha\beta}$ with the numerical coefficient of the normalization:
\begin{equation}
    \mathcal{Y}_{1s,\beta} = \left(\sqrt{\frac{9\pi}{35}}  Y_{1s,\beta}\right)^2 =
    \begin{cases}
        1.234 \quad \beta=2p\\
        1.345 \quad \beta=3p
    \end{cases} 
\end{equation}

\subsubsection{Contribution of $H_{if}^a$}

We now focus on the contribution of the initial state's antisymmetric amplitude, which contributes to the matrix element as
\begin{equation}\label{supp: NR matrix element antisymmetric term}
    H_{if}^a =  \frac{i}{\mathcal{N}_2\Lambda_{\theta}}\frac{e}{\sqrt{2\omega_k}m}\varepsilon_\lambda^{i*}c^{jk} \int_p p^k \phi^{\dagger}_\alpha(\bb p)\phi_\beta(\bb p)\int_{q} q^i q^j \,\phi^{\dagger}_\alpha(\bb q)\phi_{\alpha}(\bb q) .
\end{equation}
For the 1s level, the second integral evaluates to
\begin{equation}
    \int_{q} q^i q^j \,\phi^{\dagger}_{1s}(\bb q)\phi_{1s}(\bb q) = \frac{2mE_0}{3} \delta^{ij},
\end{equation}
which, when substituted into Eq. \eqref{supp: NR matrix element antisymmetric term}, gives
\begin{equation}\label{supp: directionality contribution}
    H_{if}^a =  \frac{i2mE_0}{3\mathcal{N}_2\Lambda_{\theta}}\frac{e}{\sqrt{2\omega_k}m}\varepsilon_\lambda^{i*}c^{ij} \int_p p^j \phi^{\dagger}_\alpha(\bb p)\phi_\beta(\bb p),
\end{equation}
where summation over the indices $i,j$ is implied. The integral does not vanish in general and, modulo constant factors, this expression differs from the undeformed matrix element $(H_{if})_0$ only by the presence of the background-field components $c^{ij}\sim\mathcal{O}(1)$ instead of $\delta^{ij}$. Therefore, we can estimate $H_{if}^a$ as follows 
\begin{equation}\label{supp: antisym element}
    H_{if}^a \sim  \frac{i2mE_0}{3\mathcal{N}_2\Lambda_\theta} (H_{if})_0.
\end{equation}
For frameworks in which time-space noncommutativity preserves unitarity and yields well-defined QFT, the components $c^{0i}$ should also contribute. However, in the dipole approximation these contributions are proportional to 
\begin{equation}
\begin{split}
    &\varepsilon^{i*}_\lambda c^{0j} \left[\int_{p} p^i  \phi_\alpha^\dagger(\bb p)\phi_\beta(\bb p) \int_q q^j \phi_\alpha^\dagger(\bb p)\phi_\alpha(\bb p) - \int_{p} p^i  \phi_\alpha^\dagger(\bb p)\phi_\alpha(\bb p) \int_q q^j \phi_\alpha^\dagger(\bb p)\phi_\beta(\bb p) \right] , \\
    &\varepsilon^{i*}_\lambda c^{j0} \left[\int_{p} p^i p^j \phi_\alpha^\dagger(\bb p)\phi_\beta(\bb p) \int_q  \phi_\alpha^\dagger(\bb p)\phi_\alpha(\bb p) - \int_{p} p^i p^j  \phi_\alpha^\dagger(\bb p)\phi_\alpha(\bb p) \int_q  \phi_\alpha^\dagger(\bb p)\phi_\beta(\bb p) \right] ,
\end{split}
\end{equation}
which vanish upon angular integration for transitions $\beta\to\alpha$ between states of opposite parity, such as 2p $\to$ 1s. Consequently, in atoms with sufficiently low $Z$, the dominant contribution is given by \eqref{supp: directionality contribution}.

\subsection{Remaining scenarios: $0<\kappa<2$ and $2<\kappa<4$}

For $0<\kappa<2$, the leading contribution arises from the symmetric component $\psi_s$ of the initial state. The corresponding matrix element is
\begin{equation}\label{supp: matrix element kappa<2}
     H_{if}^\pm =  \frac{2}{\mathcal{N}_\kappa\Lambda_{\theta}^{\kappa/2} }\frac{e}{\sqrt{2\omega_k}}\bb\varepsilon_\lambda^*\cdot \int_{p,q} \psi'^{(0)\dagger}_s(\bb p-\bb k,\bb q) \bb\alpha_{(1)} \psi_s^{(0)}(\bb p,\bb q) \sigma^{\kappa}(p-k,q)  .
\end{equation}
In contrast, for $2<\kappa<4$, the antisymmetric component $\psi_a$ provides the leading contribution and the matrix element becomes
\begin{equation}\label{supp: matrix element 2<kappa<4}
     H_{if}^\pm =  \pm\frac{2i}{\mathcal{N}_\kappa\Lambda_\theta^{2-\kappa/2}}\frac{e}{\sqrt{2\omega_k}}\bb\varepsilon_\lambda^*\cdot \int_{p,q} \psi'^{(0)\dagger}_s(\bb p-\bb k,\bb q) \bb\alpha_{(1)} \psi_a^{(0)}(\bb p,\bb q)\, (p-k) \wedge_c q .
\end{equation}

In both cases, the PEP-violating transition rates can be cast in the form
\begin{equation}\label{supp: phenomenological rate kappa<4}
    \frac{d\Gamma_\text{PV}^\pm}{d\Omega} = \left(\frac{\mathcal E}{\Lambda_\theta}\right)^{h(\kappa)} \frac{d\Gamma_0}{d\Omega} ,
\end{equation}
where the exponent $h(\kappa)$ is defined as 
\begin{equation}
    h(\kappa) = 
    \begin{cases}
        \kappa      & \kappa \in (0,2)\\
        4-\kappa &  \kappa \in(2,4)
    \end{cases}
\end{equation}
and ranges between 0 and 2. For $\kappa=1$ and $\kappa=3$, the rates are suppressed by factors of order $\mathcal{O}(\Lambda_\theta^{-1})$. The expression of the function $\mathcal{E}$ depends on the value of $\kappa$  as well as either $\sigma_\text{nr}(p,q)$ or $c^{\mu\nu}$. For $\kappa\in(0,2)$ it is given by $\mathcal{E}\sim m^{1-a/2}E_0^{a/2}$, while for $\kappa\in(2,4)$ it is $\mathcal{E}\sim \sqrt{2m E_0}$, up to order-one multiplicative coefficients.

\subsection{Directionality in PEP-violating transitions}

We now consider a radiative capture process in which an ionized helium atom captures a free electron with momentum $\bb p_0$ into a PEP-forbidden state. We focus in particular on the contribution
\begin{equation}
     H_{if}^a \sim  \bb\varepsilon_\lambda^*\cdot \int_{p,q} \psi'^{(0)\dagger}_s(\bb p-\bb k,\bb q) \bb\alpha_{(1)} \psi_a^{(0)}(\bb p, \bb q) (p-k)^ic^{ij}q^j  ,
\end{equation}
which is present in all matrix elements and for all values of $\kappa$. The antisymmetric component of the initial state reads
\begin{equation}
    \psi_a^{(0)}(\bb p, \bb q) = \frac{1}{2}[\chi(\bb p) \varphi^{(0)}_\alpha(\bb q) - \varphi^{(0)}_\alpha(\bb p)\chi(\bb q)] ,
\end{equation}
where $\alpha$ labels the state of the bound electron. For simplicity, we take the incoming electron to be a plane wave $\chi(\bb p) = u(\bb p) \delta^{(3)}(\bb p - \bb p_0)$. In the nonrelativistic limit 
\begin{equation}
    H_{if}^a \sim  \bb\varepsilon_\lambda^*\cdot \int_{p,q} \bb p \,\phi^{\dagger}_\alpha(\bb p-\bb k)\phi^{\dagger}_\alpha(\bb q) \left[ \xi_s\delta^{(3)}(\bb p-\bb p_0) \phi_{\alpha}(\bb q) - \phi_{\alpha}(\bb p) \xi_s\delta^{(3)}(\bb q -\bb p_0) \right](p-k)\wedge_c q  .
\end{equation}
The first term vanishes upon integration over $\bb q$, since
\begin{equation}\label{supp: vanishing integral}
    \int d\bb{q} |\phi_\alpha(\bb q)|^2 q^i \propto \int d\theta d\varphi \sin\theta  |\Omega_\alpha(\hat{\bb{q}})|^2 \hat{q}^i = 0.
\end{equation}
The matrix element therefore reduces to 
\begin{equation}
\begin{split}
    H_{if}^a &\sim  \bb\varepsilon_\lambda^*\cdot \int_{p,q} \bb p \,\phi^{\dagger}_\alpha(\bb p-\bb k)\phi_{\alpha}(\bb p) \phi^{*}_{\alpha|s}(\bb q) \delta^{(3)}(\bb q-\bb p_0) (p-k)\wedge_c q  \\
    &= \varepsilon_\lambda^{i*} c^{jk} \int_p p^i (p^j-k^j) \,\phi^{\dagger}_\alpha(\bb p-\bb k)\phi_{\alpha}(\bb p) \int_q q^k \phi^{*}_{\alpha|s}(\bb q)\delta^{(3)}(\bb q-\bb p_0)  \\
    &= \varepsilon_\lambda^{i*} c^{jk} p_0^k [ \Phi^{ij}(\bb k) - \Xi^{i}(\bb k)k^j ]\phi^{*}_{\alpha|s}(\bb p_0) ,
\end{split}
\end{equation}
where $\phi^{*}_{\alpha|s}(\bb q) = \phi^{\dagger}_\alpha(\bb q) \xi_s$, and we have defined 
\begin{equation}
\begin{split}
    &\Phi^{ij}(\bb k) \coloneqq \int_p p^i p^j \,\phi^{\dagger}_\alpha(\bb p-\bb k)\phi_{\alpha}(\bb p) , \\
    &\Xi^{i}(\bb k) \coloneqq \int_p p^i \,\phi^{\dagger}_\alpha(\bb p-\bb k)\phi_{\alpha}(\bb p) .
\end{split}
\end{equation}
Introducing the constant vector $\bb c = (c^{23},c^{31},c^{12})$, the expression can be rewritten as
\begin{equation}
    H_{if}^a \sim \varepsilon_\lambda^{i*} c^{jk} p_0^k \Phi^{ij}(\bb k)\phi^{*}_{\alpha|s}(\bb p_0)  - [\bb\varepsilon_\lambda^{*}\cdot\bb\Xi(\bb k)] [\bb{c} \cdot (\bb{k} \times \bb{p}_0)] \phi^{*}_{\alpha|s}(\bb p_0) .
\end{equation}
In experiments involving a large ensemble of atoms, $\Phi^{ij}$ and $\Xi^i$ average to quantities that depend only on the photon energy. The residual angular dependence is therefore governed by
\begin{equation}
    \bb{c} \cdot (\bb{k} \times \bb{p}_0),
\end{equation}
which couples the direction of the emitted photon $\hat{\bb k}$ to the background coefficients. This induces a characteristic anisotropy in PEP-violating transition rates.

\putbib

\end{bibunit}

\end{document}